\documentclass[journal,twoside]{IEEEtran}
\usepackage{graphicx,cite,amssymb,amsmath,bm}
\usepackage{mathrsfs}
\usepackage{color,soul}
\usepackage{tabulary}
\usepackage{multirow}
\usepackage{cases}
\usepackage{stfloats}
\usepackage{url}
\usepackage{booktabs}
\usepackage{algpseudocode}
\usepackage{algorithm,algpseudocode}
\usepackage{algorithmicx}

\usepackage{amsmath,amssymb}
\usepackage{mathtools} 
\usepackage{xspace}
\newcommand{\Tableref}[1]{Table~\ref{#1}}

\newcommand{\Eqref}[1]{(\ref{#1})}

\newcommand{\Secref}[1]{Section~\ref{#1}}
\newcommand{\Figref}[1]{Fig.~\ref{#1}}

\newcommand{\Thmref}[1]{Theorem~\ref{#1}}

\newcommand{\diag}{\text{diag}}

\newcommand{\herm}{^\text{H}}

\newcommand{\trans}{^\text{T}}
\newcommand{\inv}{^{-1}}

\newcommand{\bx}{\mathbf{x}}
\newcommand{\bz}{\mathbf{z}}

\newcommand{\bs}{\mathbf{s}}

\newcommand{\bP}{\mathbf{P}}

\newcommand{\bH}{\mathbf{H}}
\newcommand{\bh}{\mathbf{h}}

\newcommand{\bE}{\mathbf{E}}
\newcommand{\bg}{\mathbf{g}}

\newcommand{\bw}{\mathbf{w}}
\newcommand{\bY}{\mathbf{Y}}

\newcommand{\bI}{\mathbf{I}}

\newcommand{\bB}{\mathbf{B}}

\newcommand{\be}{\mathbf{e}}

\newcommand{\bU}{\mathbf{U}}
\newcommand{\bu}{\mathbf{u}}

\newcommand{\bN}{\mathbf{N}}

\newcommand{\bPhi}{\mathbf{\Phi}}
\newcommand{\bphi}{\boldsymbol{\phi}}

\newcommand{\bzero}{\boldsymbol{0}}
\newcommand{\mC}{\mathbb{C}}

\newcommand{\CN}{\mathcal{CN}}
\newcommand{\mrt}{\textsf{MRT}\xspace}
\newcommand{\pmrt}{\textsf{PMRT}\xspace}
\newcommand{\fzf}{\textsf{FZF}\xspace}
\newcommand{\pzf}{\textsf{PZF}\xspace}
\newcommand{\rzf}{\textsf{RZF}\xspace}
\newcommand{\ppzf}{\textsf{PPZF}\xspace}

\newcommand{\zf}{\textsf{ZF}\xspace}

\newcommand{\SNR}{\textsc{snr}\xspace}

\newcommand{\tc}{\tau_{\textsc{c}}}

\newcommand{\boundellipse}[3]
{(#1) ellipse [x radius=#2,y radius=#3]
}

\newtheorem{theorem}{Theorem}

\newtheorem{remark}{Remark}

\DeclareMathOperator{\E}{\mathsf{E}}

\newcommand{\EX}[1]{\mathsf{E}\left\{{#1}\right\}}

\newcommand{\CG}[2]{\mathcal{CN}\left({#1},{#2}\right)}

\newcommand{\C}{\mathbb{C}}
\newcommand{\cP}{\mathcal{P}}
\newcommand{\cR}{\mathcal{R}}

\newcommand{\norm}[1]{{ \left\Vert #1 \right\Vert }}

\newcommand{\setS}{\mathcal{S}}
\newcommand{\setW}{\mathcal{W}}
\newcommand{\setZ}{\mathcal{Z}}
\newcommand{\setM}{\mathcal{M}}

\newcommand{\ik}{i_{k}}

\newcommand{\tp}{\tau_{\textsc{p}}}
\newcommand{\td}{\tau_{\textsc{d}}}

\newcommand{\tsl}{\tau_{\setS_l}}
\newcommand{\tsp}{\tau_{\setS_p}}

\newcommand{\bESl}{\bE_{\setS_l}}
\newcommand{\bepsilon}{\bm{\varepsilon}}
\newcommand{\Rho}{\mathrm{\bP}}

\makeatletter
\def\@setsize#1#2#3#4{
    \@nomath#1
    \let\@currsize#1
    \baselineskip #2
    \baselineskip \baselinestretch\baselineskip
    \parskip \baselinestretch\parskip
    \setbox\strutbox \hbox{
        \vrule height.7\baselineskip
            depth.3\baselineskip
            width\z@}
    \skip\footins \baselinestretch\skip\footins
    \normalbaselineskip\baselineskip#3#4}
\makeatother

\makeatletter
\newcommand{\setstretch}[1]{
    \def\baselinestretch{#1}%
    \@currsize
    }
\makeatother



\ifCLASSOPTIONcompsoc
    \usepackage[caption=false, font=normalsize, labelfont=sf, textfont=sf, subrefformat=parens, labelformat=parens]{subfig}
\else
\usepackage[caption=false, font=footnotesize, subrefformat=parens, labelformat=parens]{subfig}
\fi

\def\BibTeX{{\rm B\kern-.05em{\sc i\kern-.025em b}\kern-.08em
    T\kern-.1667em\lower.7ex\hbox{E}\kern-.125emX}}

\newcounter{eqcnt1}
\newcounter{eqcnt2}
\newcounter{eqcnt3}
\newcounter{eqcnt4}
\newcounter{eqcnt5}
\newcounter{eqcnt6}

\begin{document}
\begin{figure*}[t!]
\normalsize
This paper was submitted for publication in IEEE Transactions on Wireless Communications on August 15, 2019.  It was finally accepted for publication on March 31, 2020. DOI: 10.1109/TWC.2020.2987027.

\

\textcopyright~2020 IEEE.  Personal use of this material is permitted.  Permission from IEEE must be obtained for all other uses, in any current or future media, including reprinting/republishing this material for advertising or promotional purposes, creating new collective works, for resale or redistribution to servers or lists, or reuse of any copyrighted component of this work in other works.
\vspace{20cm}
\end{figure*}

\newpage

\title{Local Partial Zero-Forcing Precoding for \\ Cell-Free Massive MIMO}
\author{Giovanni~Interdonato,~\IEEEmembership{Student Member,~IEEE,} Marcus Karlsson, Emil~Bj\"{o}rnson,~\IEEEmembership{Senior Member,~IEEE} and~Erik~G.~Larsson,~\IEEEmembership{Fellow,~IEEE}%
\thanks{This paper was supported in part by the European Union's Horizon 2020 research and innovation programme under grant agreement No 641985 (5Gwireless), in part by the Swedish Research Council (VR), and in part by ELLIIT. Part of this work was presented at the 2018 IEEE Global Conference on Signal and Information Processing (GlobalSIP), and conducted when G.~Interdonato was with Ericsson Research, Ericsson AB, Link\"{o}ping, Sweden.}%
\thanks{G.~Interdonato, E.~Bj\"{o}rnson, and E.~G.~Larsson are with the Department of Electrical Engineering (ISY), Link\"{o}ping University, Link\"{o}ping, Sweden (e-mail: \{giovanni.interdonato, emil.bjornson, erik.g.larsson\}@liu.se).

M.~Karlsson was with the Department of Electrical Engineering (ISY), Link\"{o}ping University, Link\"{o}ping, Sweden.}
}

\markboth{}%
{Interdonato \MakeLowercase{\textit{et al.}}: Local Partial Zero-Forcing Precoding for Cell-Free Massive MIMO}

\maketitle

\begin{abstract}
Cell-free Massive MIMO (multiple-input multiple-output) is a promising distributed network architecture for 5G-and-beyond systems. It guarantees ubiquitous coverage at high spectral efficiency (SE) by leveraging signal co-processing at multiple access points (APs), aggressive spatial user multiplexing and extraordinary macro-diversity gain.

In this study, we propose two distributed precoding schemes, referred to as \textit{local partial zero-forcing} (PZF) and \textit{local protective partial zero-forcing} (PPZF), that further improve the spectral efficiency by providing an adaptable trade-off between interference cancelation and boosting of the desired signal, with no additional front-hauling overhead, and implementable by APs with very few antennas. 

We derive closed-form expressions for the achievable SE under the assumption of independent Rayleigh fading channel, channel estimation error and pilot contamination. PZF and PPZF can substantially outperform maximum ratio transmission and zero-forcing, and their performance is comparable to that achieved by regularized zero-forcing (RZF), which is a benchmark in the downlink. Importantly, these closed-form expressions can be employed to devise optimal (long-term) power control strategies that are also suitable for RZF, whose closed-form expression for the SE is not available.

\end{abstract}

\begin{IEEEkeywords}
\noindent Cell-free Massive MIMO, distributed Massive MIMO, partial zero-forcing, precoding schemes, spectral efficiency, max-min fairness power control.
\end{IEEEkeywords}

\section{Introduction} \label{Sec:Introduction}
\IEEEPARstart{T}{he} necessity to support the incessant data traffic growth as well as to guarantee ubiquitous communication service led to deploying increasingly dense cellular networks---namely reducing the cell radius and increasing the number of antennas per site---with particular focus and effort towards how to mitigate the resulting increased inter-cell interference~\cite{Lopez2011a, Andrews2017a}.
 
Distributed communication systems~\cite{Zhou2003a} leverage signal co-processing at multiple network access points (APs) to achieve coherent combining, manage interference, provide macro-diversity gain and, as a result, achieve higher spectral efficiency. Distributed antenna systems (DAS)~\cite{Choi2007a}, Network MIMO (multiple-input multiple-output)~\cite{Venkatesan2007a}, CoMP-JT (coordinated multipoint with joint transmission)~\cite{irmer2011coordinated}, and multi-cell MIMO cooperative networks~\cite{Gesbert2010a} are some embodiments of cellular distributed systems based on such joint coherent transmission/reception from multiple APs. Ideally, inter-cell interference can be suppressed by designing cooperation clusters (i.e., the union of neighbouring cells) in which the APs jointly coherently serve all the UEs in the joint coverage area.
Early studies promised excellent theoretical gains but under the assumption of network-wide CSI knowledge at the APs. 
Network-wide coordination requires huge amount of signaling, and channel state information (CSI) to be exchanged among the APs, which poses performance limitations and system scalability issues~\cite{Lozano2013a}.
Moreover, the interference management issue was simply shifted to another level, namely from the cells to the clusters. Either inter-cell or out-of-cluster interference are inherent to the cellular paradigm and become the major bottleneck (``\textit{cooperation cannot in general change an interference-limited network to a noise-limited one}''~\cite{Lozano2013a}) as long as the network operation relies on a cell-centric (network-centric) approach. Indeed, the 3GPP LTE (3rd Generation Partnership Project Long Term Evolution) standardization of CoMP-JT has not achieved practical benefits~\cite{Fantini2016a}. 

With the advent of the Massive MIMO technology~\cite{Marzetta2016a}, suppressing both intra-cell and inter-cell interference became easier by exploiting the aggressive spatial user multiplexing deriving from the use of very large number of phase-coherently transmitting/receiving antennas at the base station and by adopting simple, linear signal processing schemes. Moreover, Massive MIMO in time-division duplex (TDD) mode allows to reduce the channel estimation overhead by leveraging the \textit{channel reciprocity}. These features have revitalized the interest towards distributed deployments~\cite{Truong2013} leading to different Massive-MIMO-based solutions. 

A recent concept, referred to as \textit{cell-free} Massive MIMO, has been introduced in~\cite{Ngo2017b}. It essentially consists in a TDD distributed Massive MIMO system wherein (potentially) all the APs coherently serve all the users (UEs) in the same time-frequency resources. Each AP is connected to a central processing unit (CPU), through a front-haul network, while the CPUs are connected through a back-haul network and are responsible for the coordination. Cell-free Massive MIMO combines the extraordinary macro-diversity from distributing many APs and the interference cancelation from cellular Massive MIMO. In a nutshell, cell-free Massive MIMO is to network MIMO as Massive MIMO is to multi-user MIMO~\cite{Interdonato2018a}. The key features that make this network infrastructure attractive are the following: $i)$ channel estimation and precoding are performed locally at each AP by leveraging the channel reciprocity and with no instantaneous CSI sharing, renouncing to a network-wide joint interference cancelation in favour of a reduced front-hauling overhead~\cite{Bjornson2010c}; $(ii)$ \textit{user-centric} perspective: each user is surrounded by serving APs and experiences no cell boundaries, hence the term cell-free. To confine the data processing within handful APs, user-specific AP subsets (possibly overlapped) can be designed on-demand (in the literature also known as dynamic cooperation clusters~\cite{Bjornson2013d}, cover-shifts~\cite{Jungnickel2014b}, and user-centric communications~\cite{Buzzi2019c}); $iii)$ predictable performance supported by rigorous spectral efficiency (SE) analysis comprising channel estimation errors and interference from pilot contamination. Moreover, optimal power control strategies, solely dependent on long-term channel statistics, can be designed based on accurate closed-form expressions for the ergodic SE~\cite{Ngo2017b}.

\subsection{Motivation}
While a comprehensive study on the uplink (UL) SE of cell-free Massive MIMO considering different levels of receiver cooperation has been recently carried out in~\cite{Bjornson2019a}, in this paper, we investigate the downlink (DL) SE provided by precoding schemes that can be implemented in a fully distributed and scalable fashion and therefore do not require any instantaneous CSI exchange between the APs and the CPUs.

Maximum ratio transmission (MRT), also known as conjugate beamforming, was advocated in~\cite{Ngo2017b} to preserve the system scalability and to cope with single-antenna APs. In~\cite{Nayebi2017a} the performance of centralized zero-forcing (ZF) processing was analyzed, but under the assumption of no pilot contamination. 
Importantly, centralized ZF requires the CPU to collect the instantaneous CSI from all its APs, construct the precoding vectors, and feed them back. Such an approach might be unscalable when the number of APs and users is large.

In our preliminary work~\cite{Interdonato2018b}, we evaluated the performance of cell-free Massive MIMO with multi-antenna APs and local \textit{full-pilot} ZF (introduced by~\cite{Bjornson2016a} for multi-cell co-located Massive MIMO). 
Each AP uses its own local channel estimates to construct a ZF precoder by which it suppresses its own interference, but not interference from other APs---unlike centralized ZF with global CSI knowledge at the CPU. 
The performance of FZF is affected by the quality of the CSI the AP is able to acquire, the available spatial degrees of freedom, i.e., the number of AP antennas $M$ and the number of orthogonal spatial directions we wish to cancel the interference towards---which is equal to the number of mutually orthogonal pilots $\tp$. 
If the condition $M > \tp$ is not fulfilled, then full-pilot ZF cannot be implemented\cite{Bjornson2016a,Interdonato2018b}. While reducing $\tp$ increases the pilot contamination and further degrades the channel estimates, larger $M$ increases in general the complexity at the AP (hardware, data processing, etc.). 

The scope of this paper is to propose a fully distributed and versatile precoding scheme providing an adaptable trade-off between interference mitigation and power increase, whose effective operation is not constrained by the number of AP antennas.

\subsection{Contributions}
The main technical contributions of the paper can be summarized as follows:
\begin{itemize}
\item We propose two fully distributed precoding schemes, referred to as \textit{local partial zero-forcing} (PZF) and \textit{local protective partial zero-forcing} (PPZF), that provide interference cancelation gain with no additional front-hauling overhead, and can be implemented by APs with very few antenna elements. 
\item For the proposed precoding schemes, we derive closed-form expressions for an achievable downlink spectral efficiency, under the assumption of independent Rayleigh fading. These expressions take into consideration channel estimation errors and pilot contamination.
\item We devise an algorithm to globally solve the max-min fairness power control optimization problem subject to per-AP power constraint. This optimization problem has structural similarity to that in~\cite{Ngo2017b} and it is based on the closed-form SE expressions we derived.
\item We quantitatively compare the performance of PZF and PPZF with MRT, FZF and local regularized ZF both with optimal max-min fairness power control and a distributed heuristic channel-dependent power control strategy.    
\end{itemize}

\section{System Model} 
\label{Sec:SysModel}

Let us consider a cell-free Massive MIMO system operating in TDD, wherein $L$ APs equipped with $M$ antennas each are able to jointly coherently serve $K$ single-antenna users, and $LM \gg K$. 
The APs are connected to multiple central processing units (CPUs) through a front-haul network. 
The channel vector between an AP $l$ and a UE (user equipment) $k$ is denoted by $\bh_{l,k} \in \C^{M \times 1}$, and captures the effects of small-scale and large-scale fading. We assume the following:
\begin{itemize}
\item perfect channel reciprocity. The channel is reciprocal as a result of a perfect calibration of the hardware chains (accurate calibration is achievable by off-the-shelf methods~\cite{Vieira2017a});
\item block-fading channel model, i.e., the channel is constant within a time-frequency interval referred to as the \textit{coherence interval}, and varies independently between coherence intervals;
\item independent Rayleigh fading, $\bh_{l,k} \sim \CN(\bzero,\beta_{l,k} \bI_M)$,
where $\beta_{l,k}$ is the large-scale fading coefficient (channel variance) between AP $l$ and UE $k$, constant over the antenna elements (i.e., $\beta_{l,k}$ does not depend on the antenna element index $m$, $m=1,\ldots,M$).
\item large-scale fading coefficients known a-priori at each AP. The large-scale fading coefficients vary slowly, in the range of several coherence intervals and depending on the UE mobility. Hence, we assume that the channel variances are estimated at an early stage and these estimates are used afterwards to estimate the current channel response;
\item infinite capacity of the front-haul network. The performance of cell-free Massive MIMO with front-haul capacity constraints was investigated in~\cite{Bashar2019a} and~\cite{Bashar2019b}.
\end{itemize}
Let $\tc$ denote the length, in samples, of the TDD frame, chosen to fit the shortest coherence interval of the users in the network. 
The TDD frame consists of three phases: $(i)$ UL pilot transmission (or UL training); $(ii)$ UL data transmission; and $(iii)$ DL data transmission. 
We let $\tp$ denote the UL training length, and $\td = \tc - \tp$ the number of samples per TDD frame spent on data transmission. 
The number of samples spent for DL data transmission and UL data transmission are given by $\xi\td$ and $(1-\xi)\td$, respectively, where $0 < \xi < 1$.

\subsection{Uplink Training}
\label{subsec:ULtraining}

{\color{black}In the UL training phase, each UE sends simultaneously a $\tp$-length pilot sequence to all the APs.}
{\color{black}Let $i_k\in \{1,\dots,\tp\}$ be the index of the pilot used by UE $k$, which is denoted by $\bphi_{i_k} \in \mC^{\tp\times1}$.} 
We assume $\tp \leq K$, meaning that some UEs can share the same orthogonal pilot sequence. In this regard, we define $\cP_k \subset\{1,\ldots,K\}$ as the set of indices, including $k$, of UEs assigned with the same pilot as UE $k$. Hence, for any UE $t$ with $t\neq k$, $\ik = i_t \Leftrightarrow t \in \cP_{k}$. The pilot sequences are mutually orthogonal and normalized such that
\begin{align*}
\bphi_{i_t}\herm \bphi_{i_k} = \begin{cases}
	0, & t \notin \cP_k, \\
	\tp, & t \in \cP_k.
	\end{cases}
\end{align*}     
The pilot signal sent by UE $k$ to all the APs is $\sqrt{p_k}\bphi_{i_k}$, where $p_k$ is the UL normalized transmit power. We assume that the UEs transmit with full power. The pilot signal received at an AP $l$ is 
\begin{equation}\label{eq:Yl}
 \bY_l = \sum\nolimits_{k=1}^{K}\bh_{l,k}\sqrt{p_k}\bphi_{i_k}\herm + \bN_{l} \in \mC^{M\times\tp},
\end{equation}
where $\bN_{l} \in \mC^{M\times\tp}$ is a Gaussian noise matrix whose elements are i.i.d. $\CG{0}{1}$. 

In order to estimate the channel to UE $k$, AP $l$ first correlates the received pilot signal with the corresponding pilot sequences $\bphi_{\ik}$, then it performs minimum mean square error estimation (MMSE). The MMSE channel estimate $\hat{\bh}_{l,k}$ can be derived as~\cite{Kay1993a}
\begin{equation} \label{eq:channel-estimate}
\hat{\bh}_{l,k} \triangleq c_{l,k}\bY_{l} \bphi_{\ik},
\end{equation}    
where $c_{l,k}$ is defined as
\begin{equation}
c_{l,k} \triangleq  \dfrac{\sqrt{p_k}\beta_{l,k}}{\tp \sum_{t\in\cP_k}p_{t}\beta_{l,t} + 1}.
\end{equation}
The estimation error is given by  $\tilde{\bh}_{l,k} = {\bh}_{l,k} - \hat{\bh}_{l,k}$. The estimate and estimation error are independent and distributed as $\hat{\bh}_{l,k}\sim\CN(\bzero, \gamma_{l,k}\bI_{M})$, and $\tilde{\bh}_{l,k}\sim\CN(\bzero, (\beta_{l,k}-\gamma_{l,k})\bI_{M})$, respectively, where $\gamma_{l,k}$ is the mean-square of the estimate, i.e., for any antenna element $m$, $m=1,\ldots,M$,
\begin{equation}
\gamma_{l,k} \triangleq \EX{|[\hat{\bh}_{l,k}]_m|^2} = \dfrac{p_k\tp\beta_{l,k}^{2}}{\tp\sum_{t\in\cP_k}p_t\beta_{l,t} + 1}.
\end{equation} 
\begin{remark}[Pilot contamination]
\label{rem:pilot-contamination}
For any pair of UEs $k$ and $t$, with $t \in \cP_k$, $t \neq k$, the respective channel estimates to any AP $l$ are linearly dependent as  
\begin{equation} \label{eq:parallel-estimates}
\hat{\bh}_{l,k} = \dfrac{\sqrt{p_k}\beta_{l,k}}{\sqrt{p_t}\beta_{l,t}}\hat{\bh}_{l,t} \quad \Rightarrow \quad \gamma_{l,k} = \frac{p_k\beta^2_{l,k}}{p_t\beta^2_{l,t}} \gamma_{l,t}. 
\end{equation}
The APs are not able to spatially separate linearly dependent channels. This is the essence of the pilot contamination.
\end{remark}

\subsection{Downlink Data Transmission}
In canonical cell-free Massive MIMO, all the APs serve all the UEs in the network, in the same time-frequency resources.
The data signal transmitted by AP $l$ to all the UEs is given by 
\begin{equation}
\bx_l = \sum\nolimits_{k=1}^{K} \sqrt{\rho_{l,k}} \mathbf{w}_{l,\ik} q_k,
\end{equation}
where $\bw_{l,\ik} \in \mC^{M \times 1}$ is the precoding vector used by AP $l$ towards UE $k$ and all the UEs using pilot $i_k$ (as result of Remark~\ref{rem:pilot-contamination}), with $\EX{\norm{\mathbf{w}_{l,\ik}}^2}=1$; $\rho_{l,k}$ is the normalized transmit power, satisfying a per-AP power constraint (described in~\Secref{sec:power-control}). The data symbol $q_k$ has unit power, $\EX{|q_{k}|^2}=1$, and zero mean, and we assume that the data symbols are uncorrelated, i.e., $\EX{q_kq_t^\ast} = 0$ for any $t \neq k$. 

\begin{remark}[Clustering]
By constraining the power control coefficient $\rho_{l,k}$ to be zero, AP $l$ is excluded from the service of UE $k$. In such a way, one can design APs cooperation clusters that serve a given UE, or, from the AP viewpoint, UEs clusters served by a given AP. 
\end{remark}

As outlined in Remark~\ref{rem:pilot-contamination}, when $\tp < K$, some of the estimated channels are parallel. Hence, the matrix of the channel estimates, $\hat{\bH}_l = [\hat{ \bh }_{l,1}, \ldots, \hat{\bh}_{l,K}]~\in~\mC^{ M \times K}$, is rank-deficient. The corresponding full-rank matrix of the channel estimates, $\bar{\bH}_l$, is given by
\begin{equation}
\bar{\bH}_l = \bY_{l}\bPhi \in \mC^{M\times \tp},
\end{equation}
where $\bPhi\!=\![\bphi_1,\ldots,\bphi_{\tp}]\!\in\!\mC^{\tp\!\times\!\tp}$ is the pilot-book matrix.
Hence, the channel estimate between AP $l$ and UE $k$ can be expressed in terms of $\bar{\bH}_l$ as
\begin{equation} \label{eq:hat-to-bar-h}
\hat{\bh}_{l,k} = c_{l,k}\bar{\bH}_l\be_{\ik},
\end{equation} 
where $\be_{\ik}$ denotes the $i_k$-th column of $\bI_{\tp}$.
\begin{remark}[Pilot-to-Precoder Mapping]
The precoders at AP $l$ are designed by using the full-rank matrix, $\bar{\bH}_l$, whose dimension is $M \times \tp$. Each AP can effectively construct $\tp$ precoding vectors, namely one per orthogonal pilot, and the same precoding vector is adopted towards those UEs sharing the same pilot. In order to construct $\bar{\bH}_l$, AP $l$ needs to acquire at least one channel estimate per uplink pilot.   
\end{remark}

At UE $k$, the received data signal can be written as
\begin{equation} \label{eq:Downlink-Signal}
\begin{split}
& y_k 
 \!=\!\sum_{l = 1}^{L} \sqrt{ \rho_{l,k} } \mathbf{h}_{l,k}\herm \mathbf{w}_{l,\ik} q_k \! + \! \sum\limits_{l=1}^{L} \sum_{t \neq k}^{K} \sqrt{ \rho_{l,t} } \mathbf{h}_{l,k}\herm \mathbf{w}_{l,i_t} q_t \! + \! n_k, 
\end{split}
\end{equation}
where the first term is the desired signal, the second term describes the multi-user interference (all the signal components intended for UE $t$, $t \neq k$), and the third term is i.i.d. Gaussian noise at the receiver, $n_k \sim\CG{0}{1}$.

\section{Performance Analysis} 
\label{sec:performance-analysis}
We evaluate the performance, in terms of downlink spectral efficiency (SE) [bit/s/Hz/user], provided by a cell-free Massive MIMO system, modeled as described in Section~\ref{Sec:SysModel}, for different precoding schemes. All the schemes considered and proposed in this section can be implemented in a distributed fashion at each AP by using only local CSI.  

\subsection{Downlink Spectral Efficiency} \label{subsec:downlinkSE}

A lower bound on the ergodic capacity, i.e., an achievable SE, can be found by using the bounding technique in~\cite[Sec. 2.3.2]{Marzetta2016a}, \cite{Medard2000a}, and~\cite{SIG-093}, known as \textit{hardening bound}. 
To obtain the hardening bound we first rewrite~\eqref{eq:Downlink-Signal} as  
\begin{align}
\label{eq:signalUE}
y_k = \text{CP}_k \cdot q_k + \text{PU}_k \cdot q_k + \sum_{t \neq k}^{K} \text{UI}_{kt} \cdot q_t + n_k,  
\end{align}  
where $\text{CP}_k$, $\text{PU}_k$, and $\text{UI}_{kt}$ represents the coherent precoding gain, precoding gain uncertainty, and multi-user interference, respectively, defined as 
\begin{align}
& \text{CP}_k = \sum\nolimits_{l=1}^L \sqrt{\rho_{l,k}} \E\left\{\mathbf{h}_{l,k}\herm \mathbf{w}_{l,\ik}\right\}, \label{eq:CB}\\
& \text{PU}_k = \sum\nolimits_{l=1}^L \left( \sqrt{\rho_{l,k}} \mathbf{h}_{l,k}\herm \mathbf{w}_{l,\ik} \! - \! \sqrt{\rho_{l,k}} \E\left\{\mathbf{h}_{l,k}\herm \mathbf{w}_{l,\ik}\right\}\right), \label{eq:BU}\\
& \text{UI}_{kt} = \sum\nolimits_{l=1}^{L} \sqrt{ \rho_{l,t} } \mathbf{h}_{l,k}\herm \mathbf{w}_{l,i_t}.
\label{eq:UI}
\end{align} 
As described in~\Eqref{eq:signalUE}, UE $k$ effectively sees a deterministic channel ($\text{CP}_k$) with some unknown noise. Since $q_k$ and $q_t$ are uncorrelated for any $t\neq k$, the first term in~\eqref{eq:signalUE} is uncorrelated with the third term. Furthermore, $q_k$ is independent of $\text{PU}_k$, thus the first and the second terms are uncorrelated. By assumption, the noise (fourth) term is independent of the first term in~\Eqref{eq:signalUE}. Therefore, the sum of the second, third, and fourth term in~\eqref{eq:signalUE} can be treated as an uncorrelated effective noise. By invoking the arguments from \cite[Sec. 2.3.2]{Marzetta2016a}, \cite{Medard2000a}, an achievable downlink SE for UE $k$ can be written as stated in~\Thmref{Theorem-Lower-Bound-Rate}.
\begin{theorem} \label{Theorem-Lower-Bound-Rate}
A lower bound on the downlink ergodic capacity{\footnote{{\color{black}Equation~\eqref{eq:DL-SE} gives a rigorous lower bound on the Shannon capacity, assuming infinitely long codewords.  In many cases rather short codewords may be used and then~\eqref{eq:DL-SE} gives overoptimistic predictions of the rate.
 
Results in finite-block length information theory are available that give lower (and upper) bounds on the transmission spectral efficiency for given codeword length, error probability and SNR, and for the AWGN channel, and these bounds are rather straightforward to obtain~\cite{Polyanskiy2010a}.  It is also known that in massive MIMO, by virtue of the channel hardening effect~\cite[Section 1.3]{Marzetta2016a}, the effective scalar channel seen by each UE is fairly close to an AWGN channel. In particular, codes optimized for the AWGN channel work also well for the channels effectively seen in massive MIMO. Hence, we expect that the relative gap between performance with infinite blocks and finite blocks in our setup is comparable to that in the AWGN channel, which in turn can be evaluated by using the theory in~\cite{Polyanskiy2010a}.}}} of a UE $k$ is given by
\begin{equation} \label{eq:DL-SE}
\mathsf{SE}_k = \xi \left( 1 - \frac{\tp}{\tc} \right) \log_2 (1 + \mathsf{SINR}_k ) \quad \text{[bit/s/Hz]},
\end{equation}
where the effective SINR is
given by~\eqref{eq:sinr:dl} at the top of this page.

Expression~\eqref{eq:DL-SE} is valid regardless of the precoding scheme used. Next, we derive closed-form expressions for the achievable SE provided by different precoding schemes, under the assumption of independent Rayleigh fading channel.

\end{theorem}

\begin{figure*}[!t]
\normalsize
\setcounter{eqcnt1}{\value{equation}}
\setcounter{equation}{14}
\begin{equation} \label{eq:sinr:dl} 
\mathsf{SINR}_k = \frac{|\text{CP}_k|^2}{\E\{|\text{PU}_k|^2\}+\sum\limits_{t \neq k}^{K} \E\{|\text{UI}_{kt}|^2\}+1} = \frac{  \left| \sum\limits_{l=1}^L \sqrt{\rho_{l,k}} \E\{ \bh_{l,k}\herm \bw_{l,\ik} \} \right|^2  }{ \sum\limits_{t=1}^K \E\left\{ \left| \sum\limits_{l=1}^L \sqrt{\rho_{l,t}} \bh_{l,k}\herm \bw_{l,i_t} \right|^2 \right\} - \left| \sum\limits_{l=1}^L \sqrt{\rho_{l,k}} \E\{ \bh_{l,k}\herm \bw_{l,\ik} \} \right|^2 + 1 }
\end{equation}
\setcounter{equation}{\value{eqcnt1}}
\hrulefill
\vspace*{4pt}
\end{figure*}

\subsection{Maximum Ratio Transmission}
For the sake of self-containment, we herein present the closed-form expression for the achievable downlink SE, if the multi-antenna APs utilize MRT. Such expression was already given in~\cite{Ngo2018a}.
The MRT precoding vector constructed by AP $l$ towards UE $k$, denoted by $\bw^\mrt_{l,i_k}$, is \addtocounter{equation}{1}
\begin{equation} \label{eq:mrt}
\bw^\mrt_{l,i_k}\! =\!  \frac{ \bar{\bH}_{l} \be_{i_k} }
{\sqrt{ \EX{ \|\bar{\bH}_{l} \be_{i_k}\|^{2}} }} \!=\! \frac{c_{l,k} \bar{\bH}_{l} \be_{i_k}}{\sqrt{\EX{\norm{\hat{\bh}_{l,k}}^2}}} \!=\! \frac{\bar{\bH}_{l} \be_{i_k}}{\sqrt{M \theta_{l,k}}},
\end{equation}
where $\theta_{l,k} = \EX{|[\bar{\bH}_l \be_{i_k}]_m|^2}$, for any antenna element $m$ of AP $l$, and $\theta_{l,k} = \gamma_{l,k}/c^2_{l,k}$.
By plugging~\eqref{eq:mrt} into~\eqref{eq:sinr:dl}, and calculating the corresponding expected values, the achievable downlink SE is obtained in closed form for MRT precoding as in Theorem~\ref{Theorem-Lower-Bound-Rate}, where
\begin{equation} \label{eq:SINR-MRT}
\mathsf{SINR}^\mrt_k \!=\! \frac{M \left( \sum\limits_{l=1}^L \sqrt{ \rho_{l,k} \gamma_{l,k}} \right)^2 }{M \!\sum\limits_{t \in \cP_k \setminus \{k\} } \!\left(\sum\limits_{l=1}^{L}\sqrt{\rho_{l,t}\gamma_{l,k}}\right)^{2}\!\!\!+\! \sum\limits_{l=1}^L \sum\limits_{t=1}^K \rho_{l,t} \beta_{l,k} \!+\! 1}.
\end{equation}

\subsection{Full-pilot Zero-Forcing Precoding}

Full-pilot zero-forcing (FZF) precoding was investigated in~\cite{Bjornson2016a} for multi-cell co-located Massive MIMO systems. Unlike canonical ZF that only suppresses intra-cell interference, FZF has also the ability to suppress inter-cell interference in a fully distributed and scalable fashion.

In cell-free Massive MIMO, FZF can be implemented by multi-antenna APs and its performance has been evaluated in our preliminary work~\cite{Interdonato2018b}. The local nature of this precoding strategy is extremely important to preserve the system scalability, which is crucial in cell-free Massive MIMO, thus we use the terminology \textit{local} FZF to stress this aspect. The local FZF precoding vector, used by AP $l$ towards UE $k$, is 
\begin{equation}  \label{eq:fzf}
\bw^{\fzf}_{l,\ik}  = \frac{ \bar{\bH}_{l} \left(\bar{\bH}_l\herm\bar{\bH}_l\right)\inv \be_{\ik} }
{\sqrt{ \E\{ \|\bar{\bH}_{l} \left(\bar{\bH}_l\herm\bar{\bH}_l\right)\inv \be_{\ik}\|^{2} \} }}.
\end{equation}
Under the assumption of independent Rayleigh fading channel, the normalization term in~\eqref{eq:fzf} is given in closed form by
\begin{align} \label{eq:den-ZF-definition}
\E\left\{ \norm{\bar{\bH}_{l} \left(\bar{\bH}_l\herm\bar{\bH}_l\right)\inv \be_{\ik}}^{2}\right\} &= \E\left\{\left[\left(\bar{\bH}_l\herm\bar{\bH}_l\right)\inv\right]_{\ik,\ik}\right\} \nonumber \\
&= \frac{1}{(M-\tp)\theta_{l,k}},
\end{align}
which follows from~\cite[Lemma 2.10]{Tulino2004}, for a $\tp \times \tp$ central complex Wishart matrix with $M$ degrees of freedom satisfying $M \geq \tp+1$\footnote{\color{black}This condition becomes $M > K$ if and only if all the UEs are assigned mutually orthogonal pilots, for which $\tp = K$ (i.e., no pilot reuse).}. We stress that, any AP $l$ can design the ZF precoders by only using its local CSI, i.e., $\bar{\bH}_l$. This yields at least two benefits:
\begin{enumerate}
\item there is no need for any centralized computation of the precoding vectors at the CPU. As a consequence, there is neither exchange of instantaneous CSI from the APs to the CPU nor information about the computed precoding vectors fed back from the CPU to the APs.
\item lower complexity. The precoding vector design requires the computation of the pseudo-inverse matrix $\bar{\bH}_l\left(\bar{\bH}_l\herm\bar{\bH}_l\right)\inv$, where $\bar{\bH}_l$ has dimension $M \times \tp$. In contrast, canonical ZF, performed at the CPU, would require global CSI knowledge and the computation of the pseudo-inverse matrix $\bar{\bH}\left(\bar{\bH}\herm\bar{\bH}\right)\inv,$ where $\bar{\bH}\!\!~=~\!\!\left[ \bar{\bH}_1, \ldots, \bar{\bH}_L \right]\trans$ has dimension $LM \times \tp$.    
\end{enumerate}
FZF suppresses interference towards all the UEs unless they share the same pilot:
\begin{align} \label{eq:ZF-Property}
\alpha^\zf_{l,k,t} &\triangleq \hat{\bh}_{l,k}\herm \bw^{\zf}_{l,i_t} \nonumber \\
&= \left(c_{l,k}\bar{\bH}_{l}\be_{i_k}\right)\herm\bar{\bH}_{l} \left(\bar{\bH}_{l}\herm\bar{\bH}_{l}\right)\inv\be_{i_t} \sqrt{\theta_{l,t}(M-\tp)} \nonumber \\
&=
c_{l,k}\be_{i_k}\herm\be_{i_t}\sqrt{(M-\tp)\theta_{l,t}} \nonumber \\
&=
\begin{cases}  0, &  t \notin \cP_k, \\
\sqrt{(M-\tp)\gamma_{l,k}}, & t \in \cP_k. \end{cases}
\end{align}
Importantly, with local FZF, an AP can only suppress its own interference, but not interference from other APs (as instead global zero-forcing would be able to do). The capability to cancel interference is highly dependent on the quality of the acquired CSI, and on the number of AP's antennas, which must meet the requirement $M > \tp$.
 
By substituting~\eqref{eq:ZF-Property} into~\eqref{eq:sinr:dl}, and computing the corresponding expected values, the ergodic SE is obtained in closed form for full-pilot ZF precoding as in Theorem~\ref{Theorem-Lower-Bound-Rate}, where
the effective SINR is
given by~\eqref{eq:SINR-ZF-alternative} at the top of this page.
\begin{IEEEproof}
See Appendix A.
\end{IEEEproof}

Unlike centralized ZF, local FZF is a scalable precoding scheme as it is implemented in a fully distributed fashion. Moreover, the latter has lower complexity and allows faster precoder computation due to the smaller pseudo-inverse matrices.  
Compared to MRT, local FZF provides interference cancelation gain with no additional front-hauling overhead.
The cost is a loss in array gain of $\tp$.

\begin{figure*}[!t]
\normalsize
\setcounter{eqcnt2}{\value{equation}}
\setcounter{equation}{20}
\begin{equation} \label{eq:SINR-ZF-alternative}
\mathsf{SINR}_k^{\fzf} = \frac{ (M-\tp) \left( \sum\limits_{l =1}^L \sqrt{ \rho_{l,k} \gamma_{l,k}} \right)^2 }{ (M-\tp) \sum\limits_{t \in \cP_k \setminus \{k\} } \left(\sum\limits_{l =1}^{L}\sqrt{\rho_{l,t}\gamma_{l,k}}\right)^{2} + \sum\limits_{l=1}^L \sum\limits_{t=1}^K \rho_{l,t} \left( \beta_{l,k} - \gamma_{l,k} \right)+ 1 }
\end{equation}
\setcounter{equation}{\value{eqcnt2}}
\hrulefill
\vspace*{4pt}
\end{figure*}

\subsection{Local Partial Zero-Forcing Precoding} \label{subsec:partialZF}
In this section, we describe our first proposed precoding scheme named \textit{local ``partial'' zero-forcing} (PZF). The principle behind this scheme is that each AP only suppresses the interference it causes to the \textit{strongest} UEs, namely the UEs with the largest channel gain and that it presumably interferes the most with. Conversely, the interference caused to the \textit{weakest} UEs is tolerated. More specifically, for any AP $l$, the set of the active UEs is virtually divided in two disjoint subsets: $(i)$ strong UEs, and $(ii)$ weak UEs. 
We define $\setS_l \subset \{1,\ldots,K\}$, and $\setW_l \subset \{1,\ldots,K\}$  as the set of indices of strong and weak UEs, respectively. Note that $\setS_l \cap \setW_l = \varnothing$, $|\setS_l| + |\setW_l| = K$.  
The UE grouping can follow different criteria. For instance, it may be based on the mean-square of the channel gain: a UE $k$ belongs to $\setS_l$ if $\beta_{l,k}$ is above a predetermined threshold, else UE $k$ belongs to $\setW_l$. 
\begin{remark}[Grouping co-pilot UEs]
\label{rem:grouping}
UEs assigned with the same pilot are grouped together as an AP is not able to separate them spatially. Let UE $t \in \cP_k$ and UE $k \in \setS_l$, then UE $t \in \setS_l$.
\end{remark}

Let $\tsl \leq \tp$ be the number of different pilots used by the UEs $\in \setS_l$, and $\cR_{\setS_l}\!=\!\left\{r_{l,1}, \ldots, r_{l,\tsl} \right\}$ the set of the corresponding pilot indices. The matrix that collects only the pilots of the UEs~$\in \setS_l$ is given by 
$\bPhi_{\setS_l} = \bPhi \bESl$, where $\bESl\!\!\!\!~=~\!\!\![ \be_{r_{l,1}}, \ldots, \be_{r_{l,\tsl}}] \in \mC^{\tp \times \tsl}$, and $\be_{r_{l,i}}$ is the $r_{l,i}$-th column of $\bI_{\tp}$.
Let $j_{l,k} \in \left\{1,\ldots,\tsl \right\}$ be the index, with respect to $\bPhi_{\setS_l}$, of the pilot used by UE $k \in \setS_l$. We define $\bepsilon_{j_{l,k}} \in \mC^{\tsl}$ as the $j_{l,k}$-th column of $\bI_{\tsl}$, which yields $\bESl \bepsilon_{j_{l,k}} = \be_{\ik}$.

Local PZF operates as follows: AP $l$ transmits to all the UEs~$\in \setS_l$ by using local FZF, and to all the UEs~$\in \setW_l$ by using MRT. The signal $\bx_l$, sent by AP $l$ employing PZF, is thus given by \addtocounter{equation}{1}
\begin{equation} \label{eq:pzf-tx}
\bx^{\pzf}_l = \sum_{k \in \setS_l} \sqrt{\rho_{l,k}} \bw^\pzf_{l,\ik} q_k + \sum_{j \in \setW_l} \sqrt{\rho_{l,j}} \bw^\mrt_{l,i_j} q_j,
\end{equation}
where $\bw^\mrt_{l,i_j}$ is given in~\eqref{eq:mrt}, and $\bw^\pzf_{l,\ik}$ is defined as
\begin{equation} \label{eq:pZF}
\bw^\pzf_{l,\ik}  = \frac{ \bar{\bH}_{l} \bESl \left(\bESl\herm\bar{\bH}_l\herm\bar{\bH}_l\bESl\right)\inv \bepsilon_{j_{l,k}} }
{\sqrt{ \E\left\{\norm{\bar{\bH}_{l}\bESl \left(\bESl\herm\bar{\bH}_l\herm\bar{\bH}_l\bESl\right)\inv \bepsilon_{j_{l,k}}}^{2} \right\}}}.
\end{equation}
Under the assumption of independent Rayleigh fading channel, the normalization term in~\eqref{eq:pZF} is given, in closed form, by
\begin{equation} \label{eq:den-pZF}
\E\left\{\!\norm{\bar{\bH}_{l}\bESl \left(\bESl\herm\bar{\bH}_l\herm\bar{\bH}_l\bESl\right)\inv \! \bepsilon_{j_{l,k}}}^{2}\!\right\}\!=\!\frac{1}{(M\!-\!\tsl)\theta_{l,k}},
\end{equation}
which follows from~\cite[Lemma 2.10]{Tulino2004}, for a $\tsl \times \tsl$ central complex Wishart matrix with $M$ degrees of freedom satisfying $M \geq \tsl + 1$. 
\begin{remark}[PZF generalizes FZF]
The PZF precoding vector reduces to the FZF precoding vector if $\setS_l=\{1,\ldots,K\}$ (or equivalently $\setW_l = \varnothing),~\forall l, l=1,\ldots,L$. In fact, if $\setS_l=\{1,\ldots,K\}$, then $\tsl = \tp$ and $\bESl = \bI_{\tp}$. As a result, $\bepsilon_{j_{l,k}} = \be_{\ik}$ and~\eqref{eq:pZF} becomes identical to~\eqref{eq:fzf}.  
\end{remark}

PZF only orthogonalizes the $\tsl$ channels in the matrix $\bar{\bH}_{l}\bESl$. The (\textit{intra-group}) interference between UEs $\in \setS_l$ is actively suppressed, while the (\textit{inter-group}) interference between UEs $\in \setS_l$ and UEs $\in \setW_l$ is managed as in MRT. Hence, for any pair of UEs $k, t \in \setS_l$ 
\begin{align} \label{eq:PZF-Property}
\alpha^\pzf_{l,k,t}  \triangleq \hat{\bh}_{l,k}\herm \bw^\pzf_{l,i_t}
&= 
\begin{cases}  0, &  t \notin \cP_k, \\
\sqrt{(M-\tsl)\gamma_{l,k}}, & t \in \cP_k, \end{cases}
\end{align} 
which is computed by following the same approach in~\eqref{eq:ZF-Property}.
For any pair of UEs $k, t \in \setW_l$
\begin{align} \label{eq:PZF-MRT-Property}
\EX{\hat{\bh}_{l,k}\herm \bw^\mrt_{l,i_t}} &= 
\begin{cases}  0, &  t \notin \cP_k, \\
\sqrt{M \gamma_{l,k}}, & t \in \cP_k. \end{cases}
\end{align}
If $t \notin \cP_k$, the expectation in~\eqref{eq:PZF-MRT-Property} is zero, since $\hat{\bh}_{l,k}$ is independent of  $\bw^\mrt_{l,i_t}$ and zero-mean RV. If $t \in \cP_k$, then $\bw^\mrt_{l,i_t}$ and $\hat{\bh}_{l,k}$ are linearly dependent and the result in~\eqref{eq:PZF-MRT-Property} is obtained by applying~\eqref{eq:mrt}. For any pair of UEs $k, t$ in different groups, it must hold that $t \notin \cP_k$, since co-pilot UEs are placed in the same group. Hence, $\hat{\bh}_{l,k}$ is independent of $\bw_{l,i_t}$, and $\EX{\hat{\bh}_{l,k}\herm \bw_{l,i_t}} = 0$.

Clearly, PZF is performed locally at each AP, and it does not require any CSI to be exchanged between APs and CPU. The UE grouping varies from AP to AP as it is based on local statistical CSI, but neighbouring APs might have similar or identical sets $\setS$ and $\setW$.
From the UE viewpoint, a UE $k$ is differently served by two disjoint subsets of APs. Let $\setZ_k$ and $\setM_k$ denote the set of indices of APs that transmit to UE $k$ by using $\bw^\pzf_{l,\ik}$ and $\bw^\mrt_{l,\ik}$, respectively, defined as 
\begin{align*}
\setZ_k &\triangleq \{ l : k \in \setS_l, l=1,...,L \}, \\
\setM_k &\triangleq \{ l : k \in \setW_l, l=1,...,L \},
\end{align*}
with $\setZ_k \cap \setM_k = \varnothing$, $|\setZ_k| + |\setM_k| = L$.      
The received data signal at the UE $k$ can be written as
\begin{align}
y_k \!&=\! \left(\sum\limits_{l \in \setZ_k} \sqrt{ \rho_{l,k} } \bh_{l,k}\herm \bw^\pzf_{l,\ik} \!+\! \sum\limits_{p \in \setM_k}\!\sqrt{ \rho_{p,k} } \bh_{p,k}\herm \bw^\mrt_{p,\ik} \right) q_k \nonumber \\
&\quad+ \sum\limits_{ \substack{t = 1 \\ t \neq k} }^K \! \left( \sum\limits_{l \in \setZ_t} \sqrt{ \rho_{l,t} } \bh_{l,k}\herm \bw^\pzf_{l,i_t}\!+\!\sum\limits_{p \in \setM_t}\!\sqrt{ \rho_{p,t} } \bh_{p,k}\herm \bw^\mrt_{p,i_t}\! \right) q_t \nonumber \\
&\quad+ n_k.
\end{align}
Utilizing the same capacity bounding technique as in Section~\ref{subsec:downlinkSE}, the effective SINR, assuming PZF precoding, is given by~\eqref{eq:sinr-pzf} at the top of this page.
\begin{figure*}[!t]
\normalsize
\setcounter{eqcnt3}{\value{equation}}
\setcounter{equation}{27}
\begin{equation} \label{eq:sinr-pzf}
\mathsf{SINR}^{\pzf}_k \!= \frac{  \left| \sum\limits_{l \in \setZ_k} \sqrt{\rho_{l,k}} \E\{ \bh_{l,k}\herm \bw^\pzf_{l,\ik} \} + \sum\limits_{p \in \setM_k} \sqrt{\rho_{p,k}} \E\{ \bh_{p,k}\herm \bw^\mrt_{p,\ik} \} \right|^2  }{ \sum\limits_{t=1}^K \E\left\{ \left| \sum\limits_{l \in \setZ_t}\!\! \sqrt{\rho_{l,t}} \bh_{l,k}\herm \bw^\pzf_{l,i_t} \!+\!\!\! \sum\limits_{p \in \setM_t}\!\! \sqrt{\rho_{p,t}} \bh_{p,k}\herm \bw^\mrt_{p,i_t}  \right|^2 \right\} \!-\! \left| \sum\limits_{l \in \setZ_k} \sqrt{\rho_{l,k}} \E\{ \bh_{l,k}\herm \bw^\pzf_{l,\ik} \} \!+\!\!\! \sum\limits_{p \in \setM_k}\!\! \sqrt{\rho_{p,k}} \E\{ \bh_{p,k}\herm \bw^\mrt_{p,\ik} \} \right|^2 \!+\! 1 }
\end{equation}
\setcounter{equation}{\value{eqcnt3}}
\hrulefill
\vspace*{4pt}
\end{figure*}
By plugging~\eqref{eq:pZF} and~\eqref{eq:mrt} into~\eqref{eq:sinr-pzf}, and computing the expected values, the ergodic SE is obtained in closed form for PZF precoding as in Theorem~\ref{Theorem-Lower-Bound-Rate}, where the SINR is given by~\eqref{eq:sinr-pzf-final} at the top of the next page.
\begin{figure*}[!t]
\normalsize
\setcounter{eqcnt4}{\value{equation}}
\setcounter{equation}{28}
\begin{equation} \label{eq:sinr-pzf-final}
\mathsf{SINR}^{\pzf}_k = \frac{  \left( \sum\limits_{l=1}^L \sqrt{(M-\delta_{l,k}\tsl)\rho_{l,k}\gamma_{l,k}}\right)^2  }{ \sum\limits_{t\in \cP_k \setminus \{k\}}\left(\sum\limits_{l=1}^L\sqrt{(M-\delta_{l,t}\tsl)\rho_{l,t}\gamma_{l,k}}\right)^{2} + \sum\limits_{t=1}^{K}\sum\limits_{l=1}^L\rho_{l,t}(\beta_{l,k}-\delta_{l,t}\delta_{l,k}\gamma_{l,k}) + 1 },
\end{equation}
where $\delta_{l,k}$ is defined as 
\begin{equation} \label{eq:delta-def}
\delta_{l,k} \triangleq
\begin{cases}
1 \quad \text{if } l \in \setZ_k, \\
0 \quad \text{if } l \in \setM_k.
\end{cases}
\end{equation}
\setcounter{equation}{\value{eqcnt4}}
\hrulefill
\vspace*{4pt}
\end{figure*}
\begin{IEEEproof}
See Appendix B.
\end{IEEEproof}
For any UE $k$, the array gain from AP $l$ is either $M$, if $k \in \setW_l$, or $M-\tsl$, if $k \in \setS_l$. The latter is larger than (at most equal to) the array gain that FZF would provide, since $\tsl \leq \tp$.
Note that, if $\setZ_k=\{1,...,L \}~\forall k, k=1,\ldots,K$, then all APs serve all the UEs with FZF, thus $\tsl = \tp$, $\delta_{l,k}=1~\forall l,k$, and~\eqref{eq:sinr-pzf-final} reduces to~\eqref{eq:SINR-ZF-alternative}. Conversely, if $\setM_k=\{1,...,L \}~\forall k, k=1,\ldots,K$, then all APs serve all the UEs with MRT, thus $\tsl = 0$, $\delta_{l,k}=0~\forall l,k$, and~\eqref{eq:sinr-pzf-final} reduces to~\eqref{eq:SINR-MRT}.

Pointing out the role of the function $\delta_{\cdot,\cdot}$, the SINR expression in~\eqref{eq:sinr-pzf-final} tells us that, for any UE $k$, the coherent precoding gain (i.e., numerator) depends on the precoding scheme used towards UE $k$, while the coherent interference (i.e., first term of the denominator) depends solely on the precoding scheme used towards the co-pilot UEs (any UE $t\in \cP_k \setminus \{k\}$). Interestingly, the non-coherent interference (i.e., second term of the denominator) can be significantly reduced only if both UE $k$ and  any UE $t$ are in $\setS_l$. Hence, all the UEs in the strong UE set suffer from, besides pilot contamination, non-coherent interference due to the UEs served by MRT. 

In the literature, the principle of using a number of antennas for interference cancelation and the rest for signal boosting is not certainly new. Different flavours of partial zero-forcing precoding/combining schemes were analyzed in the context of MIMO communications~\cite{Jindal2011,Veetil2015}, and recently in millimeter wave cellular networks~\cite{Fang2017} and cell-free Massive MIMO~\cite{Buzzi2018}. However, earlier works assume perfect CSI knowledge and the key novelty of our work is how we combine the channel estimation with the construction of the PZF precoding vector.\footnote{\color{black}Our proposed method is a generalization of the approach used in~\cite{Jindal2011}. For instance, apart from the different normalization adopted, \cite[Eq. (9)]{Jindal2011} can be obtained from~\eqref{eq:pZF} by setting $|\setS_l|=\kappa, \forall l, l = 1,\ldots, L$, for any integer $\kappa$, $1 \leq \kappa \leq K$.}

\subsection{Local Protective Partial Zero-Forcing}
As described in the previous section, with PZF, the UEs in $\setS_l$ experience, besides interference from pilot contamination, non-coherent interference from the signals transmitted to the UEs in $\setW_l$.
To significantly reduce this interference, we propose an enhanced PZF scheme that guarantees full ``protection'' to the strong UEs by forcing the MRT to take place in the orthogonal complement of $\bar{\bH}_l\bE_{\setS_l}$. We refer to this scheme as \textit{protective partial zero-forcing} (PPZF). Let \addtocounter{equation}{3}
\begin{equation}
\bB_l = \bI_M - \bar{\bH}_l\bESl\left(\bESl\herm\bar{\bH}_l\herm\bar{\bH}_l\bESl\right)\inv \bESl\herm\bar{\bH}_l\herm,
\end{equation}
denote the projection matrix onto the orthogonal complement of $\bar{\bH}_l\bE_{\setS_l}$. The MRT precoding vector from AP $l$ to the UEs $\in \setW_l$ is now given by
\begin{equation}
\label{eq:PZF-MRT}
\bw^\pmrt_{l,i_j} =  \frac{\bB_l\bar{\bH}_{l} \be_{i_j} }
{\sqrt{ \E\{ \|\bB_l\bar{\bH}_{l} \be_{i_j}\|^{2} \} }} 
= 
\frac{\bB_l\bar{\bH}_{l} \be_{i_j} }
{\sqrt{ (M-\tsl)\theta_{l,j} }}.
\end{equation}
By design, we have $\hat{\bh}_{l,k}\herm \bB_l = 0, \text{ if }k\in \setS_l$.
The effective per-user SINR achieved by this scheme is equal to~\eqref{eq:sinr-pzf}, but replacing $\bw^\mrt_{l,i_j}$ with $\bw^\pmrt_{l,i_j}$. The ergodic SE for the PPZF scheme is given in closed form by Theorem~\ref{Theorem-Lower-Bound-Rate}, where the effective SINR is given by~\eqref{eq:sinr-pzf-final:protected} at the top of the next page.
\begin{IEEEproof}
See Appendix C.
\end{IEEEproof}
Comparing~\eqref{eq:sinr-pzf-final} with~\eqref{eq:sinr-pzf-final:protected}, we observe that the use of PPZF gives the array gain $M-\tsl$ in any case, regardless of whether $k \in \setS_l$. Importantly, the non-coherent interference solely depends on the precoding scheme used towards UE $k$, meaning that if $k \in \setS_l$, then the non-coherent interference almost vanishes (a small contribution survives due to the channel estimation errors).    

The philosophy of this precoding scheme might be summarized with the motto ``to protect and to serve'', in that it guarantees full interference protection---except for pilot contamination---to UEs with good channel conditions, while still providing service to UEs with poor channel conditions. PPZF offers a balance between interference cancelation and boosting of the desired signal. This balance is adjustable by properly setting the UE grouping criterion, letting $\tsl$ satisfy the condition $M > \tsl$. Such a versatile scheme can provide excellent interference cancelation gains even with APs equipped with few antenna elements.

\subsection{Local Regularized Zero-Forcing}
Similarly to PZF and PPZF, regularized zero-forcing (RZF) offers, but simultaneously to all the UEs, a trade-off between interference suppression and boosting of the intended signal~\cite{Peel2005a}.
 
The regularized zero-forcing (RZF) precoding vector is obtained by adding a regularization term---a diagonal matrix whose elements relate to the $\SNR^{-1}$ at each UE---to the matrix to be inverted when defining the pseudo-inverse of the channel estimates. 
In this paper, we are interested in a local RZF scheme wherein the precoding vector is function only of the local channel estimates collected at each AP. The RZF precoding vector designed by AP $l$ towards UE $k$ is, similarly to~\cite{SIG-093}, defined as \addtocounter{equation}{1} 
\begin{equation}  \label{eq:rzf}
\bw^{\rzf}_{l,k}  = \frac{ \hat{\bH}_{l} \left(\hat{\bH}_l\herm\hat{\bH}_l + \Rho_l\inv \right)\inv \hat{\be}_k }
{\sqrt{ \EX{\norm{\hat{\bH}_{l} \left(\hat{\bH}_l\herm\hat{\bH}_l + \Rho_l\inv \right)\inv \hat{\be}_k}^2} }},
\end{equation}
where $\Rho_l = \diag(\rho_{l,1}, \ldots, \rho_{l,K}) \in \mathbb{R}^{K \times K}$, and $\hat{\be}_k$ is the $k$-th column of $\bI_K$. Unlike the precoding schemes described above, an AP must construct $K$ RZF precoding vectors, one for each UE, if different power levels are allocated among the UEs. 

Deriving a closed-form expression for the achievable SE is, due to the regularization term, intractable and beyond the scope of this paper. Hence, we evaluate the achievable SE by using~\eqref{eq:DL-SE}, inserting~\eqref{eq:rzf} into~\eqref{eq:sinr:dl}, and computing the corresponding expectations by Monte-Carlo simulations for any choice of the RZF precoding vectors.

\begin{figure*}[!t]
\normalsize
\setcounter{eqcnt5}{\value{equation}}
\setcounter{equation}{32}
\begin{equation}
\label{eq:sinr-pzf-final:protected}
\mathsf{SINR}^{\ppzf}_k = \frac{  \left( \sum\limits_{l=1}^L \sqrt{(M-\tsl)\rho_{l,k}\gamma_{l,k}}\right)^2  }{ \sum\limits_{t\in \cP_k \setminus \{k\}}\left(\sum\limits_{l=1}^L\sqrt{(M-\tsl)\rho_{l,t}\gamma_{l,k}}\right)^{2} + \sum\limits_{t=1}^{K}\sum\limits_{l=1}^L\rho_{l,t}(\beta_{l,k}-\delta_{l,k}\gamma_{l,k}) + 1 },
\end{equation}
\setcounter{equation}{\value{eqcnt5}}
\hrulefill
\vspace*{4pt}
\end{figure*}

\section{Power Control}
\label{sec:power-control}
\subsection{Max-Min Fairness}
\label{subsec:max-min}
Max-min fairness power control consists in maximizing the lowest user's downlink SE, and providing uniform service throughout the network.
The (normalized) transmit power at AP $l$,
\begin{equation} \label{eq:Transmit-Power}
\EX{\norm{\bx_l}^2} = \sum\limits_{k=1}^K \rho_{l,k} \EX{\norm{\bw_{l,\ik}}^2 } = \sum\limits_{k=1}^K \rho_{l,k},
\end{equation}
is constrained by $\rho_l^{\mathsf{max}}$ as
\begin{equation}
\label{eq:per-AP-power-contraint}
\sum\nolimits_{k=1}^{K} \rho_{l,k} \leq \rho_l^{\mathsf{max}}, \; \forall l.
\end{equation}
The max-min fairness power control optimization problem{\color{black}, subject to per-AP power constraints,} is formulated as follows 
\begin{equation} \label{Problem:Max-Min-QoS-Original-SINR}
\begin{aligned}
& \underset{ \{ \rho_{l,k} \geq 0 \}}{\textrm{maximize}} & & \min_k \mathsf{SINR}_k^{ \mathsf{ps} } \\
& \textrm{subject to} & & \sum\nolimits_{k=1}^{K} \rho_{l,k} \leq \rho_l^{\mathsf{max}}, \; \forall l,\\
\end{aligned}
\end{equation}
where the superscript ``$\mathsf{ps}$'' stands for ``precoding scheme'', $\mathsf{ps}=\{ \mathsf{MRT}, \mathsf{FZF}, \mathsf{PZF}, \mathsf{PPZF} \}$, and
\begin{equation} \label{eq:SINR-scheme}
\mathsf{SINR}_k^{ \mathsf{ps} } = \frac{\left(\sum\nolimits_{l=1}^L \sqrt{\rho_{l,k} g^{\mathsf{ps}}_{l,k,k}} \right)^2 }{\sum\limits_{t \in \cP_k \setminus \{k\} }\! \left(\sum\limits_{l=1}^{L}\sqrt{\rho_{l,t} g^{\mathsf{ps}}_{l,k,t}}\right)^{2}\!\!\!+\! \sum\limits_{l=1}^L \sum\limits_{t=1}^K \rho_{l,t} z^{\mathsf{ps}}_{l,k,t}\!+\! 1}.
\end{equation} 
The values that the terms $g^{\mathsf{ps}}_{l,k,t}$ and $z^{\mathsf{ps}}_{l,k,t}$ assume for the precoding schemes presented in~\Secref{sec:performance-analysis}, are summarized in~\Tableref{tab:SINR-terms}.
\begin{table}[!t]
\normalsize
\centering
\renewcommand{\arraystretch}{1}
\caption{$g^{\mathsf{ps}}_{l,k,t}$ and $z^{\mathsf{ps}}_{l,k,t}$ for the precoding schemes in~\Secref{sec:performance-analysis}.}
\begin{tabular}{c | c | c }
 & $g^{\mathsf{ps}}_{l,k,t}$ & $z^{\mathsf{ps}}_{l,k,t}$\\
\hline
$\mathsf{MRT}$ & $M \gamma_{l,k}$ & $\beta_{l,k}$\\
$\mathsf{FZF}$ & $(M-\tp) \gamma_{l,k}$ & $\beta_{l,k}-\gamma_{l,k}$\\
$\mathsf{PZF}$ & $(M-\delta_{l,t}\tsl) \gamma_{l,k}$ & $\beta_{l,k}-\delta_{l,k}\delta_{l,t}\gamma_{l,k}$\\
$\mathsf{PPZF}$ & $(M-\tsl) \gamma_{l,k}$ & $\beta_{l,k}-\delta_{l,k}\gamma_{l,k}$\\
\end{tabular}
\vspace*{-5mm}
\label{tab:SINR-terms}
\end{table}
The equivalent epigraph representation of \eqref{Problem:Max-Min-QoS-Original-SINR} is shown in~\eqref{Problem:Max-Min-QoS2} at the top of this page, where $\nu$ is the minimum SINR among the UEs that we aim to maximize. The max-min optimization problem in~\eqref{Problem:Max-Min-QoS2} has structural similarity to that in~\cite{Ngo2017b}. Following the same approach as in~\cite{Ngo2017b}, we next show that problem~\eqref{Problem:Max-Min-QoS2} can be solved for a fixed $\nu$ as a second-order cone program {\color{black}(while the UL max-min power allocation problem can be solved by using geometric programming~\cite{Ngo2017b,Bashar2019})}. 
\begin{figure*}[!t]
\normalsize
\setcounter{eqcnt6}{\value{equation}}
\setcounter{equation}{38}
\begin{equation} \label{Problem:Max-Min-QoS2}
\begin{aligned}
& \underset{ \{ \rho_{l,k} \geq 0 \},~\nu}{\textrm{maximize}}
& & \nu \\
& \textrm{subject to}
& & \frac{\left(\sum\nolimits_{l=1}^L \sqrt{\rho_{l,k} g^{\mathsf{ps}}_{l,k,k}} \right)^2 }{\sum\limits_{t \in \cP_k \setminus \{k\} } \left(\sum\limits_{l=1}^{L}\sqrt{\rho_{l,t} g^{\mathsf{ps}}_{l,k,t}}\right)^{2}+ \sum\limits_{l=1}^L \sum\limits_{t=1}^K \rho_{l,t} z^{\mathsf{ps}}_{l,k,t}+ 1} \geq \nu \;, \forall k \\
& & & \sum\nolimits_{k=1}^{K} \rho_{l,k} \leq \rho_l^{\mathsf{max}} \;, \forall l,\\
\end{aligned}
\end{equation}
\setcounter{equation}{\value{eqcnt6}}
\hrulefill
\vspace*{4pt}
\end{figure*}
Let $\bU = [\bu_1, \ldots , \bu_K ] \in \mC^{M \times K}$ have columns $\bu_{t}~=~ \left[\sqrt{\rho_{1,t}}, \ldots, \sqrt{\rho_{L,t}}\right]\trans, \text{ for } t = 1, \ldots, K$. Let $\bu_{i}'$ denote the $i$-th row of $\bU$. We also let $\bz_{t,i}= \left[\sqrt{z^{\mathsf{ps}}_{1,t,i}}, \ldots, \sqrt{z^{\mathsf{ps}}_{L,t,i}}\right]\trans $ and $\bg_{t,i} = \left[\sqrt{g^{\mathsf{ps}}_{1,t,i}}, \ldots, \sqrt{g^{\mathsf{ps}}_{L,t,i}}\right]\trans$.
Furthermore,
\begin{align*}
\bs_k &= \Big[\sqrt{\nu} \Big( \bg_{k,t'_{1}}\trans \bu_{t'_{1}}, \ldots, \bg_{k,t'_{\left|\cP_k \setminus \{k\}\right|}}\trans \bu_{t'_{\left|\cP_k \setminus \{k\}\right|}}, \Big.\Big. \nonumber \\  
&\qquad \Big. \Big. \norm{\bz_{k,1} \circ \bu_1}, \ldots, \norm{\bz_{k,K} \circ \bu_K}, 1 \Big)\Big]\trans \in \mC^{K+ |\mathcal{P}_k|},
\end{align*} where $t_{1}^{'}, \ldots, t_{|\cP_k\! \setminus \{k\} |}^{'}$ are the UE indices $\in \cP_k \setminus \{k\}$, and $\circ$ denotes the element-wise (Hadamard) product. Finally, \eqref{Problem:Max-Min-QoS2} is reformulated as \addtocounter{equation}{1}
\begin{subequations} \label{eq:SOCP} 
\begin{align}
 \underset{ \{ \rho_{i,t} \geq 0 \},~\nu}{\textrm{maximize}}
 & \quad \nu \\
 \mbox{subject to}
 & \quad \norm{\mathbf{s}_k} \leq \mathbf{g}_{k,k}\trans \mathbf{u}_k, \; \forall k,\label{subeq:sk-SOCP} \\ 
 & \quad \norm{\mathbf{u}_{i}'} \leq \sqrt{\rho_i^{\mathsf{max}}}, \; \forall i, \label{subeq:power-SOCP}
 \end{align}
\end{subequations}
The constraints~\eqref{subeq:sk-SOCP} and~\eqref{subeq:power-SOCP} are both second-order cones with respect to $\{\rho_{i,t}\}$, but jointly in $\{\rho_{i,t}\}$ and $\nu$. Consequently, \eqref{eq:SOCP} is a convex program if $\nu$ is fixed, and the optimal solution can be obtained by using interior-point methods, e.g., CVX toolbox~\cite{cvx2012}. Moreover, since~\eqref{subeq:sk-SOCP} is increasing function of $\nu$, the solution to \eqref{eq:SOCP} is obtained by solving the corresponding feasibility problem, through \textit{bisection method}~\cite{Boyd2004a}.

Optimal power control, although attractive from the performance viewpoint, poses significant implementation and computational challenges. Firstly, such a centralized power control policy requires the APs to send long-term channel statistics to the CPU, where the power control coefficients are computed and subsequently fed back. Secondly, solving the max-min optimization problem, although a convex problem, might be prohibitive for large number of APs and UEs, since the computation time is polynomial in the number of optimization variables, $LK$. Optimal power control might, therefore, undermine the system scalability, increase the front-hauling overhead and be computationally very demanding for large $L$ and $K$. On the other hand, the power control coefficients computation, which depends only on large-scale fading quantities, occurs whenever there are macroscopic network variations, typically several coherence intervals. Therefore, when the updating frequency of the power control coefficients is relatively low, optimal power control is efficiently practicable.

\subsection{Distributed Heuristic Channel-Dependent Strategy}
A less performant but scalable solution consists in implementing power control in a distributed fashion at each AP, and letting the power control coefficients depend exclusively on the local long-term channel statistics. More specifically, setting the power control coefficients to be proportional (at a proper rate) to users' channel gain yields good performance~\cite{Ngo2017b,Buzzi2019c,Riera2018,Interdonato2019a}. Similarly, in this paper, we consider a distributed heuristic power control policy where the power control coefficients are set as
\begin{equation}
\label{eq:heuristic-distributed-power-control-coefficients}
\rho_{l,k} = \frac{\gamma_{l,k}}{\sum_{i=1}^K \gamma_{l,i}} \rho_l^{\mathsf{max}}, \forall l, \forall k.
\end{equation}
{\color{black}Inserting~\eqref{eq:heuristic-distributed-power-control-coefficients} into~\eqref{eq:per-AP-power-contraint}, the latter holds with equality, meaning that the APs transmit with full power. Moreover, $\rho_{l,k} \propto \gamma_{l,k}$ implies that the better the channel (the larger $\gamma_{l,k}$) is the more power is allocated.}

\section{Front-hauling Costs and Computational Complexity}
\label{sec:fronthaul-complexity}
{\color{black}Front-hauling costs refer to the amount of information to exchange via the front-haul network needed to perform joint coherent transmission/detection and other centralized network operations. The uplink front-haul signaling load of a cell-free Massive MIMO system, expressed as the number of complex scalars to send from the APs to the CPU via the front-haul network, is investigated in~\cite{Bjornson2019a}, for different levels of cooperation between APs and CPU. In the downlink, since all precoding schemes herein considered are fully distributed, the front-hauling load concerns essentially the transmission of data payload from the CPU to the APs. Specifically, in each coherence block, the CPU needs to send at most (in the case that all the APs serve all the UEs) $\td KL$ complex scalars (i.e., $q_k$). Moreover, if the centralized optimal max-min power control, described in~\Secref{subsec:max-min}, is adopted, then the downlink front-hauling load increases. Indeed, $LK$ real-valued scalars, i.e., the optimal power control coefficients $\{\rho_{l,k}\}$ computed at the CPU, need to be sent to the APs. 
In practice, the resolution of the signal quantization may be optimized, see~\cite{Bashar2019a} and~\cite{Bashar2019b} for some initial studies on this topic.

Next, we provide rigorous expressions for the computational complexity of the proposed precoding schemes, obtained using the same methodology as in~\cite[Sections 4.1.2, 4.3.2]{SIG-093} and shown in~\Tableref{tab:complexity} at the top of this page.

\begin{table*}[!t]
\centering
\normalsize
\renewcommand{\arraystretch}{1.1}
\caption{Computational complexity per coherence block}
\begin{tabular}{c|c|c|c}
\multirow{2}{*}{
	\begin{tabular}[c]{@{}c@{}}Precoding\\ 	
		Scheme
	\end{tabular}
} & \multicolumn{2}{c|}{Computing precoding vectors} & 
\multirow{2}{*}{
	\begin{tabular}[c]{@{}c@{}}Transmission\\ 
		Multiplications
	\end{tabular}
} \\
& Multiplications & Divisions & \\
\midrule
FZF & $\dfrac{3 \tp^2 M}{2} + \dfrac{\tp M}{2} + \dfrac{\tp^3-\tp}{3}$ & $\tp$ & $\td M K$ \\ 
PZF & $\dfrac{3 \tsl^2 M}{2} + \dfrac{\tsl M}{2} + \dfrac{\tsl^3-\tsl}{3}$ & $\tsl$ & $\td M K$ \\
PPZF & $\dfrac{3 \tsl^2 M}{2} + \dfrac{\tsl M}{2} + \dfrac{\tsl^3-\tsl}{3} + 2(\tp-\tsl) \tsl M$ & $\tsl$ & $\td M K$ \\
\end{tabular}
\label{tab:complexity}
\vspace*{4pt}
\\ \hrulefill
\end{table*}
Only complex multiplications and divisions are considered, whereas additions and subtractions are neglected since these are much less complex. The average normalization factor in the precoding vector can be absorbed into the scalar signal $q_k$ and calculated over multiple coherence blocks, thus the corresponding complexity can be neglected, see~\cite[Sections 4.1.2, 4.3.2]{SIG-093} for further details. 

\Figref{fig:complexity} shows the normalized computational complexity per coherence block of the proposed precoding schemes versus $\tsl$, given $M=16$ and $\tp=10$.
\begin{figure}[!t]
\centering
\includegraphics[width=.8\linewidth]{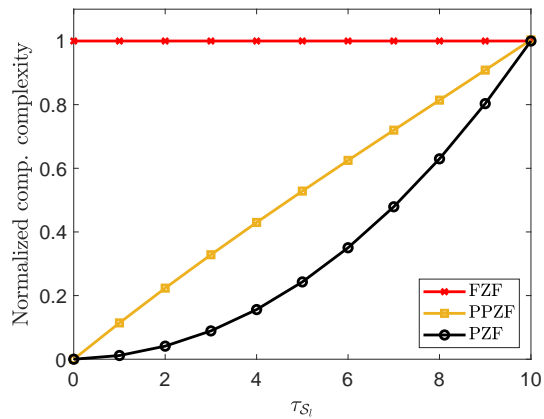}
\caption{Normalized computational complexity per coherence block of the proposed precoding schemes. $M=16$~and~$\tp=10$.}
\label{fig:complexity}
\end{figure} 
We define the computational complexity as the sum of the number of complex multiplications and divisions (shown in~\Tableref{tab:complexity}).  
The normalized computational complexity is obtained by dividing the computational complexity by the complexity of FZF. 
As shown in~\Figref{fig:complexity}, the complexity of PZF and PPZF is lower than the complexity of FZF due to the fact that $\tsl \leq \tp$. The complexity of PPZF is, as expected, higher than that of PZF as PPZF requires $2(\tp-\tsl) \tsl M$ more complex multiplications when computing the $(\tp-\tsl)$ MRT precoding vectors in~\eqref{eq:PZF-MRT}.
Lastly, the complexity of computing the $\td$ transmit signals $\sum_{k=1}^K \sqrt{\rho_{l,k}} \bw_{l,i_k} q_k$ at AP $l$ is the same for all the proposed precoding schemes and at most (if AP $l$ serves all the $K$ users) equal to $\td M K$.           
}

\section{Simulation Results} \label{sec:numerical-results}
The performance of the proposed precoding schemes are numerically evaluated, analyzed and discussed in this section. We firstly introduce the network setup and the parameters considered in our simulations.

\subsection{Simulation Scenario}
We consider an area of size $D \times D$ squared meters, wrapped-around by eight twin areas in order to reduce edge effects.
{\color{black}APs and UEs are uniformly distributed at random. The cumulative distribution function (CDF) of the SE presented next corresponds to the SE values collected over 500 network snapshots for different random realizations of the AP/UE locations. For each snapshot, lower bounds on the ergodic SE are obtained in closed form, as described in~\Secref{sec:performance-analysis}, conditioned on the large-scale fading coefficients.}    

The large-scale fading coefficients $\{\beta_{l,k}\}$ incorporate pathloss and shadow fading, as follows
\begin{equation}
\label{eq:beta}
\beta_{l,k} = \mathsf{PL}_{l,k} \cdot 10^{\frac{\sigma_\text{sh}z_{l,k}}{10}},
\end{equation}  
where $\mathsf{PL}_{l,k}$ represents the pathloss, and $10^{\frac{\sigma_\text{sh}z_{l,k}}{10}}$ models log-normal shadow fading with standard deviation $\sigma_\text{sh}$ and $z_{l,k}\sim\mathcal{N}(0,1)$. 
The pathloss follows the 3GPP Urban Microcell model in \cite[Table B.1.2.1-1]{LTE2017}, which, assuming a 2~GHz carrier frequency, is given by
\begin{equation}
\label{eq:path-loss-model}
\mathsf{PL}_{l,k}~\text{[dB]} = -30.5-36.7 \log_{10} \left(\frac{d_{l,k}}{1~\text{m}} \right),
\end{equation}
where $d_{l,k}$ is the distance between AP $l$ and UE $k$ including AP and UE's heights. 
The shadow fading accounts for spatial correlations both between APs and between UEs by 
\begin{equation}
\label{eq:shadowing-model}
z_{l,k} = \sqrt{\varrho}~a_l + \sqrt{1-\varrho}~b_k,
\end{equation}  
where $a_l \sim \mathcal{N}(0,1)$ and $b_k \sim \mathcal{N}(0,1)$ are independent RVs modeling the shadow fading impact on the channels from AP $l$ to all the UEs and from UE $k$ to all the APs, respectively, and the parameter $\varrho$ provides weighting for these impacts. The shadowing terms are correlated as
\begin{equation*}
\EX{a_l a_i} = 2^{-\frac{d^{\mathsf{AP}}_{l,i}}{9\text{ m}}}, \qquad \EX{b_k b_t} = 2^{-\frac{d^{\mathsf{UE}}_{k,t}}{9\text{ m}}},
\end{equation*}    
where $d^{\mathsf{AP}}_{l,i}$ is the distance between AP $l$ and  AP $i$, $d^{\mathsf{UE}}_{k,t}$ is the distance between UE $k$ and UE $t$, and 9 meters is the decorrelation distance~\cite{LTE2017}.  

The following settings are adopted in the simulations, unless otherwise stated: $D = 1000$ m, $\sigma_\text{sh}=4$~dB, AP height 10 m, UE height 1.5 m, channel bandwidth $B=20$ MHz. We take $\xi=0.5$ (i.e., symmetric TDD frame), $\tc = 200$ samples of the time-frequency grid corresponding to a coherence bandwidth of 200 kHz and a coherence time of 1 ms. The maximum transmit power is $200$ mW for each AP (to be divided among the antennas), and $100$ mW for each UE, while the noise power is $w^{\mathsf{(dBm)}}_\mathsf{p}=-92$ dBm. Hence,
\begin{align*}
\rho_l^{\mathsf{max}} \text{ [dBm]} &= 10\log_{10}(200) - w^{\mathsf{(dBm)}}_\mathsf{p}, \forall l, l=1,\ldots,L, \\
p_k \text{ [dBm]} &= 10\log_{10}(100) - w^{\mathsf{(dBm)}}_\mathsf{p}, \forall k, k = 1,\ldots,K.
\end{align*}  
Finally, we assume that the $\tp < K$ UL pilot sequences are randomly assigned to the UEs.

\subsection{Performance Evaluation}

\begin{figure}[!t] \centering
	\subfloat[Downlink per-user spectral efficiency.\label{fig:fig1a}]{ 
    	\includegraphics[width=.8\columnwidth]{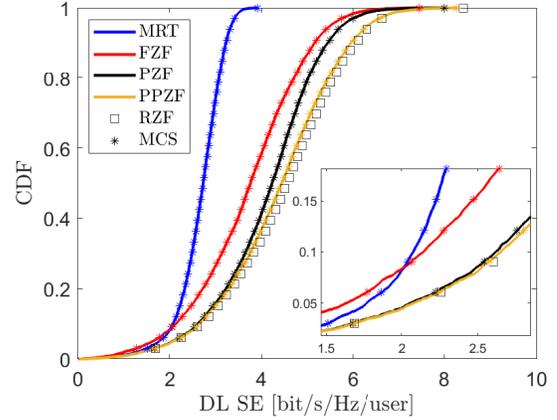}}\hfill
    \subfloat[Downlink sum spectral efficiency\label{fig:fig1b}]{
    	\includegraphics[width=.8\columnwidth]{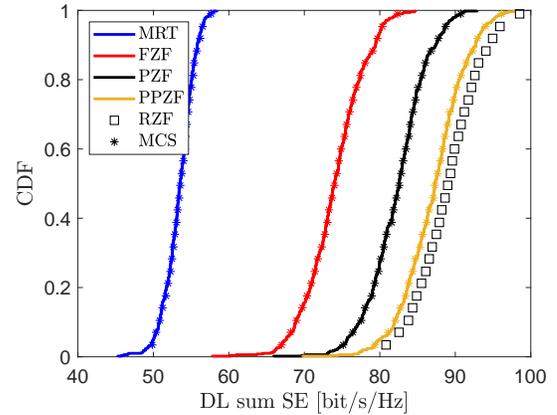}}\hfill
	\caption{CDFs of the SE achieved by different distributed precoding schemes. Simulation setup: $L=200, M=16, \tp=15$, and $K=20$. Power control is based on~\eqref{eq:heuristic-distributed-power-control-coefficients}, while the UEs grouping strategy follows~\eqref{eq:UE-splitting}. Solid curves and markers indicate the results obtained in closed form and by Monte-Carlo simulations (MCS), respectively.}
	\label{fig:fig1} 
	\vspace*{-2mm} 
\end{figure}

\Figref{fig:fig1} shows the CDFs of the SEs achieved by the precoding schemes described in~\Secref{sec:performance-analysis}. In this initial comparison, the setup consists in $L=200, M=16, \tp=15$, and $K=20$. The heuristic distributed channel-dependent power control policy (hereafter HCD) is adopted. The UEs grouping strategy in PZF and PPZF relies, inspired by~\cite{Ngo2018a}, on the following rule
\begin{equation}
\label{eq:UE-splitting}
\sum\limits_{k=1}^{|\hat{\setS_l}|} \frac{\bar{\beta}_{l,k}}{\sum\nolimits_{t=1}^K \beta_{l,t}} \geq \upsilon\%,
\end{equation}
according to which AP $l$ constructs its set $\hat{\setS_l}$ by selecting the UEs that contribute at least $\upsilon\%$ of the overall channel gain. The final strong UE set $\setS_l$ is given by the UEs in $\hat{\setS_l}$ plus the remaining UEs that use any pilot used in $\hat{\setS_l}$.
In~\eqref{eq:UE-splitting}, $\{\bar{\beta}_{l,1}, \ldots, \bar{\beta}_{lK}\}$ indicates the set of the large-scale fading coefficients sorted in descending order. In the example shown in~\Figref{fig:fig1}, we set $\upsilon=0.95$. For any AP $l$, $\upsilon$ is conveniently adjusted (lowered), if the resulting $\tsl$ does not fulfil the condition $M>\tsl$.
From Fig.~\subref*{fig:fig1a}, we first observe that {\color{black}the gain of FZF, PZF, and PPZF with respect MRT is quite significant, especially for high percentiles. The reason is two-fold: primarily due to the ability of the former to suppress the multi-user interference; secondarily because the HCD power control intrinsically boosts the SE of UEs with good channel conditions.} Compared to FZF, PZF and PPZF provide up to 25\% improvement in terms of per-user SE. In this example, FZF suffers from modest array gain $M-\tp$ as almost all the degrees of freedom are exploited to cancel the interference. {\color{black}Such a small array gain leads to poor performance, especially at low percentiles, where FZF performs slightly worse than MRT.}
By contrast, PZF and PPZF only cancel the interference among the strong UEs, using $\tsl \leq \tp$ degrees of freedom and taking advantage of a larger array gain. Hence, it is not necessary to cancel interference towards all the available orthogonal directions.
PPZF improves the upper SE percentiles compared to PZF, thanks to its protective nature towards the UEs with larger channel gain. Up to an additional 7\% can be gained, in terms of sum SE, by using PPZF, as shown in~Fig.~\subref*{fig:fig1b}.
PPZF and PZF have great ability to mitigate both coherent and non-coherent interference, while MRT and FZF suffer from excessive coherent interference but for two different reasons: MRT does not suppress interference by nature, whereas FZF experiences small array gain since all the available degrees of freedom are exploited to cancel the interference. 
Lastly, from~\Figref{fig:fig1} we can observe that the results obtained in closed form (solid curves) and by Monte-Carlo simulations (markers), for each precoding schemes, are precisely overlapped, which numerically validates our derived closed-form expressions. 

\begin{figure}[!t]
\centering
\includegraphics[width=.8\columnwidth]{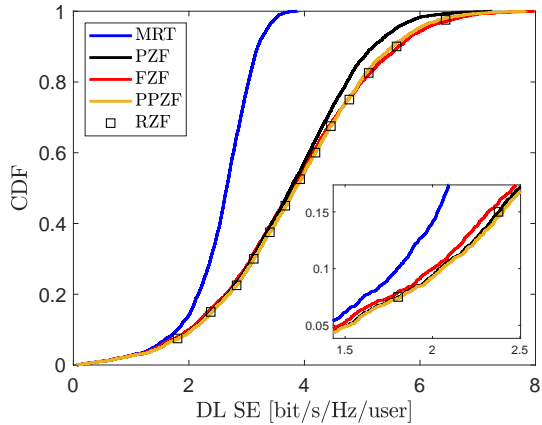}
\caption{CDFs of the per-user SE achieved by different distributed precoding schemes. Simulation settings resemble those in~\Figref{fig:fig1} with only one difference: $\tp=10$.}
\label{fig:fig2}
\vspace*{-2mm}
\end{figure}

In~\Figref{fig:fig2}, we slightly change the simulation scenario by setting $\tp=10$. By reducing $\tp$, we increase the pilot re-use, hence the pilot contamination. Consequently, the ability to suppress the interference reduces and, compared to the setup in~\Figref{fig:fig1}, PZF and PPZF perform worse. The median SE of FZF remains almost constant (about 3.8 bit/s/Hz/user) as FZF can still benefit from the increased array gain. Under these assumptions FZF, PZF and PPZF perform equally {\color{black}but significantly better than MRT, although this performance gap reduces at low percentiles due to the inegalitarian nature of the HDC power control strategy.}
Regardless of the setup, PPZF performs as well as RZF (benchmark), suggesting that~\eqref{eq:sinr-pzf-final:protected} might be a reliable closed-form expression to estimate the performance of RZF.

\begin{figure}[!t]
\centering
\includegraphics[width=\columnwidth]{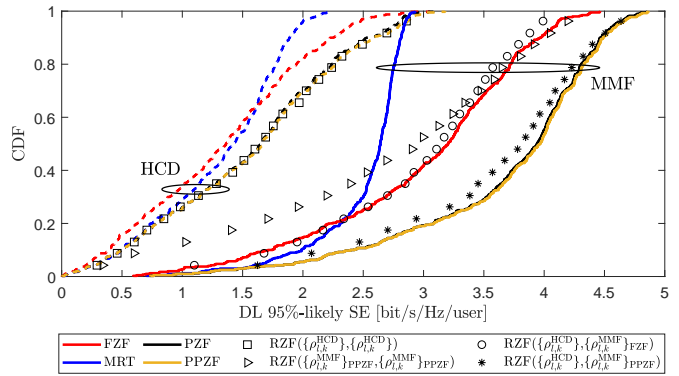}
\vspace*{-10mm}
\caption{CDFs of the per-user SE for different precoding schemes, with HCD and MMF power control. The notation RZF$\left(\{\rho^{\mathrm{HCD}}_{l,k}\},\{\rho^{\mathrm{MMF}}_{l,k}\}_{\mathrm{PPZF}} \right)$ indicates the RZF scheme where coefficients $\{\rho^{\mathrm{HCD}}_{l,k}\}$ are used to define the regularization term in~\eqref{eq:rzf}, while $\{\rho^{\mathrm{MMF}}_{l,k}\}$, resulting from~\eqref{Problem:Max-Min-QoS-Original-SINR} with $\mathsf{ps} = \mathsf{PPZF}$, are employed in~\eqref{eq:sinr:dl} for power allocation.}
\label{fig:fig3}
\vspace*{-2mm}
\end{figure}

In~\Figref{fig:fig3} we present the results achieved by using max-min fairness (MMF) power control. The simulation setup consists in $L=100, M=8, \tp=7$, and $K=10$. We keep the AP and UE density constant by reducing the simulation area to $500 \times 500$ squared meters. 
Firstly, we point out how powerful MMF is for increasing the minimum service provided throughout the network. For instance, PPZF with MMF can guarantee up to 6-fold 95\%-likely SE improvement over PPZF with HCD power control. With MMF power control, PZF and PPZF are the best precoding schemes and are identical: the opportunistic nature of PPZF is balanced out by the egalitarian philosophy of MMF. Since a closed-form expression for RZF is not available, we evaluate its performance numerically (by Monte-Carlo simulations), plugging different sets of power control coefficients into~\eqref{eq:rzf} and~\eqref{eq:sinr:dl}.      
In~\Figref{fig:fig3}, the notation RZF$\left(\{\rho^{\mathrm{HCD}}_{l,k}\},\{\rho^{\mathrm{MMF}}_{l,k}\}_{\mathrm{PPZF}} \right)$ describes the performance achieved by RZF where the power control coefficients in~\eqref{eq:heuristic-distributed-power-control-coefficients} are used in~\eqref{eq:rzf} to construct the precoding vectors, and the optimal power control coefficients resulting from~\eqref{Problem:Max-Min-QoS-Original-SINR}, with $\mathsf{ps} = \mathsf{PPZF}$, are employed in~\eqref{eq:sinr:dl} for power allocation. As we can see, the performance of RZF are comparable to our proposed schemes PZF and PPZF only if  
the coefficients $\{\rho^{\mathrm{MMF}}_{l,k}\}$ optimized for PPZF are used for power allocation and $\{\rho^{\mathrm{HCD}}_{l,k}\}$ are used to define the regularization term in~\eqref{eq:rzf}. The dynamic range of the coefficients $\{\rho^{\mathrm{HCD}}_{l,k}\}$ is much smaller than that of $\{\rho^{\mathrm{MMF}}_{l,k}\}$, thus the former are more suitable to regularize the matrix inversion.

\begin{figure}[!t] \centering
	\subfloat[$L=100, K=10, \tp=7$, and $D=500$ m.\label{fig:fig4a}]{ 
    	\includegraphics[width=.8\columnwidth]{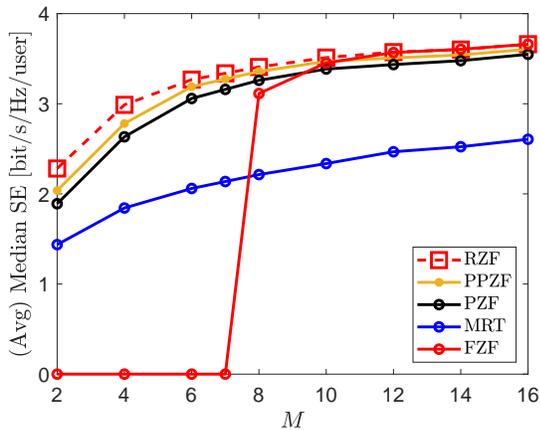}}\hfill
    \subfloat[$L=100, K=15, M=8$, and $D=500$ m.\label{fig:fig4b}]{
    	\includegraphics[width=.8\columnwidth]{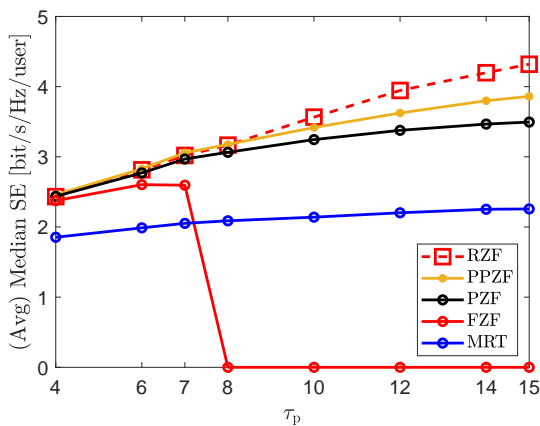}}\hfill
	\caption{Median SE, averaged over several large-scale fading realizations, achieved by different precoding schemes. Power control is based on~\eqref{eq:heuristic-distributed-power-control-coefficients}, while the UEs grouping strategy follows~\eqref{eq:UE-splitting}.}
	\label{fig:fig4}  
	\vspace*{-3mm}
\end{figure}

In~\Figref{fig:fig4} we emphasize the implementation versatility of PZF and PPZF versus the limitations of the FZF scheme. When $\tp$ is fixed, PZF, PPZF and RZF can be implemented by APs equipped with any number of antenna elements, while to implement FZF, the condition $M > \tp$ must be fulfilled---else the FZF pseudo-inverse matrix is not defined. Fig.~\subref*{fig:fig4a} shows, for instance, that 6 antennas are needed for PPZF to guarantee a median SE of 3 bit/s/Hz/user, against 8 antennas for FZF. From Fig.~\subref*{fig:fig4b}, we observe that FZF constraints the number of available orthogonal pilots, and much higher SEs can be achieved by RZF, PPZF and PZF in the operation regime in which FZF cannot be implemented.          

\begin{figure}[!t] \centering
	\subfloat[Median SE versus UE grouping threshold in~\eqref{eq:UE-splitting}.\label{fig:fig6a}]{ 
    	\includegraphics[width=.83\columnwidth]{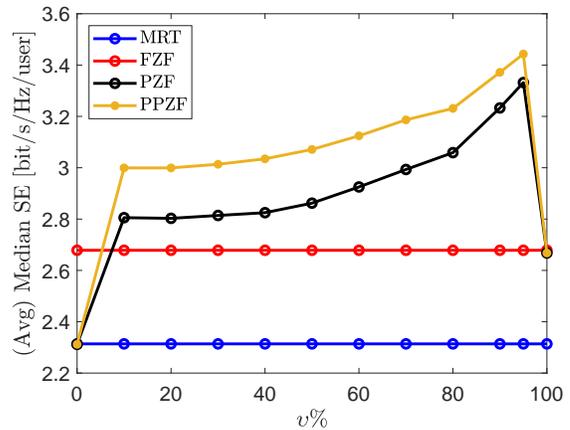}}\hfill
    \subfloat[Median SE versus AP clustering threshold. $\upsilon = 95\%$.\label{fig:fig6b}]{
    	\includegraphics[width=.81\columnwidth]{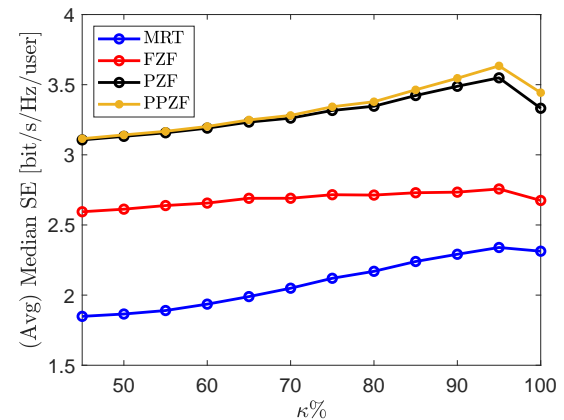}}\hfill
	\caption{Median SE, averaged over many large-scale fading realizations, for different precoding schemes. Simulation settings resemble those in~\Figref{fig:fig1}, but with $K=40$.}
	\label{fig:fig6} 
	\vspace*{-3mm} 
\end{figure}

The performance of PZF and PPZF deeply depend on the criterion used to split strong and weak UEs. Until now, in our simulations we have considered the UE grouping strategy in~\eqref{eq:UE-splitting} with $\upsilon=95\%$. 
The choice of this threshold value needs further motivation. Fig. \subref*{fig:fig6a} illustrates the median per-user SE, averaged over many large-scale fading realizations, versus $\upsilon$. Simulation settings resemble those in~\Figref{fig:fig1}, but with $K=40$. 
The optimal SE can be achieved when the strong UE set consists of the UEs whose channel gain, towards a given AP, corresponds to the 95\% of the overall channel gain. The remaining UEs are grouped in the weak UE set, unless they do not share any pilot with UEs in the strong UE set (see Remark \ref{rem:grouping}). With this setup, $1/3$--$1/2$ of the UEs are selected to be part of $\setS_l$ and $\tsl$ is 4--6, on average. A large value for $\upsilon$ (e.g., $\upsilon=100\%$) would group more UEs in the strong UEs set, and employ more degrees of freedom to suppress modest levels of interference at the cost of a reduced array gain. The loss in array gain is larger then the additional interference cancelation gain. On the other hand, a small value for $\upsilon$ would group more UEs in the weak UE set, and employ less antennas to cancel interference. However, intolerable interference would decrease the SE, despite the increased array gain.
Note that, PZF and PPZF reduce to MRT if $\upsilon=0\%$, or to FZF if $\upsilon=100\%$.

The same approach as in~\eqref{eq:UE-splitting} can be used to select a subset of APs that will serve a given UE~\cite{Ngo2018a,Interdonato2018a}. AP selection (or AP clustering) is necessary to preserve the system scalability, i.e., the ability of the network to handle a growing amount of work (data processing, signaling, power control, etc.) by adding UEs to the system. Fig. \subref*{fig:fig6b} shows the median per-user SE, averaged over many large-scale fading realizations, versus $\kappa$, i.e., the threshold used to form the user-specific AP clusters as follows
\begin{equation}
\sum\limits_{l=1}^{|\mathcal{A}_k|} \frac{\bar{\beta}_{l,k}}{\sum\nolimits_{p=1}^L \beta_{p,k}} \geq \kappa\%,
\end{equation}
where $|\mathcal{A}_k|$ is the cardinality of the user-$k$-specific AP cluster, and $\{\bar{\beta}_{1,k}, \ldots, \bar{\beta}_{L,k}\}$ is the set of the large-scale fading coefficients sorted in descending order.
The simulation settings resemble those in Fig.~\subref*{fig:fig6a} and here we set $\upsilon = 95\%$. For all the precoding schemes, it turns out that the optimal value of $\kappa$ is around $95\%$ which, in this setup, corresponds to 15--20 handful APs (out of 200) participating in the service of a given UE, on average. Changing perspective, each AP will serve a subset of UEs in the network, ignoring the remaining UEs. 
Firstly, the user-specific AP clusters are formed. Then, each AP groups its UEs into strong and weak UE sets. Since some of the UEs are dropped at the first stage, the cardinality of the weak UE set will be reasonably smaller so it is the interference due to the transmission towards the weak UEs. This explains the almost identical performance of PPZF and PZF in Fig.~\subref*{fig:fig6b}.

\begin{figure}[!t]
\centering
\includegraphics[width=.8\columnwidth]{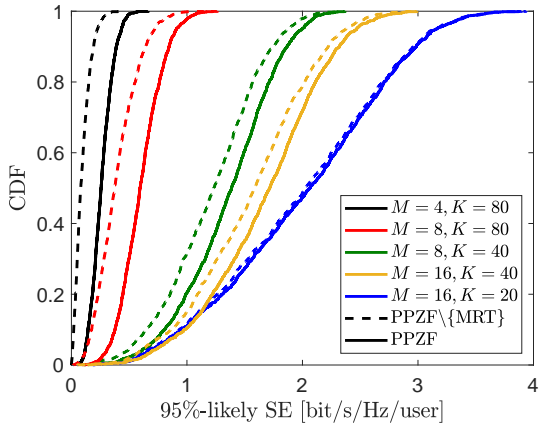}
\caption{CDF of the 95\%-likely SE achieved by PPZF and PPZF with no service towards the weak UEs (denoted by PPZF~\textbackslash $\{ \mathrm{MRT} \}$). Settings: $L~=~200, \tp=K/2,  \upsilon=95\%$, and $\kappa=95\%$. Power control is based on~\eqref{eq:heuristic-distributed-power-control-coefficients}.}
\label{fig:fig7}
\vspace*{-2mm}
\end{figure}

One may wonder whether serving the weak UEs pays off. The answer is given in~\Figref{fig:fig7} which shows the CDF of the 95\%-likely SE achieved by PPZF and the variant of PPZF, denoted by PPZF~\textbackslash $\{ \mathrm{MRT} \}$, in which the weak UEs receive no service, i.e., there is no transmission by using MRT. We focus on the 95\%-likely SE since caring of the UEs with poor channel conditions might have only impact on the lower percentiles of the CDF of the SE.   
As we can see in~\Figref{fig:fig7}, serving the weak UEs yields substantial gains when the ratio $K/M$ ($\tp/M$) is large, as UEs with relatively good channel gain might be grouped by many APs in the weak UE set due to the lack of degrees of freedom and receive no service. Conversely, if the ratio $K/M$ is small, then it is quite likely that any UE $k$ belongs to many strong UE sets, and receiving service from the APs in $\setM_k$ does not bring any additional benefit.    

\section{Conclusion}
The proposed local partial zero-forcing and local protective partial zero-forcing are  versatile distributed precoding schemes that can significantly improve the spectral efficiency of a cell-free Massive MIMO system, compared to the traditional MRT and ZF precoding schemes. Especially, protective partial zero-forcing may, in many scenarios of practical interest, outperform partial zero-forcing by providing full interference protection towards the users with better channel conditions. The proposed schemes perform as well as regularized zero-forcing (benchmark), and the corresponding closed-form expressions we derived are valuable to devise optimal power control strategies that are also suitable for regularized zero-forcing. Moreover, regularized zero-forcing requires the APs to construct one precoding vector per user, while one precoding vector per orthogonal pilot is needed when partial zero-forcing or protective partial zero-forcing are implemented. This results in lower complexity as $\tp \leq K$.

{\color{black}The combination of cell-free massive MIMO and millimeter wave (mmWave) communications is appealing as cell-free massive MIMO, given the extraordinary macro-diversity provided by the joint coherent distributed transmission, can significantly mitigate the severe path, penetration and diffraction losses characterizing the mmWave propagation. Investigating how the proposed precoding schemes perform in such a LoS-dominated sparse-scattering environment, and how they cope with blockage effects, certainly constitutes an attractive future research direction, especially considering the promising results, in terms of uplink coverage probability, given~in~\cite{Fang2017}.}

\section*{Appendix}
\subsection{Proof of Corollary 1}
We first compute in closed form the numerator in~\eqref{eq:sinr:dl} as
\begin{align} \label{eq:fzf-num}
&\left| \sum\limits_{l=1}^L\!\sqrt{\rho_{l,k}} \EX{ \bh_{l,k}\herm \bw^\zf_{l,\ik}} \right|^2 \!\!\! \nonumber \\
&\quad=\!\! \left(\sum\limits_{l=1}^{L} \sqrt{\rho_{l,k}} \alpha^\zf_{l,k,k} \right)^{2}\!\!\!=\! (M\!-\!\tp)\!\left(\sum\limits_{l=1}^{L}\sqrt{\rho_{l,t}\gamma_{l,k}}\right)^{2}\!\!, 
\end{align}
where $\alpha^\zf_{l,k,k}$ is defined in~\eqref{eq:ZF-Property}. The first term in the denominator of~\eqref{eq:sinr:dl} is given by
\begin{align} \label{eq:fzf-den}
&\sum\limits_{t=1}^{K} \E\left\{\left|\sum\limits_{l=1}^{L}\sqrt{\rho_{l,t}}\bh_{l,k}\herm \bw^\zf_{l,i_t}\right|^{2}\right\} \nonumber \\
&= \sum\limits_{t=1}^{K} \E\left\{\left|\sum\limits_{l=1}^{L}\sqrt{\rho_{l,t}}(\alpha^\zf_{l,k,t} + \tilde{\bh}_{l,k}\herm \bw^\zf_{l,i_t})\right|^{2}\right\} \nonumber \\
&\stackrel{(a)}{=} \sum\limits_{t=1}^{K} \sum\limits_{l=1}^{L}\sum\limits_{l'=1}^{L} \sqrt{\rho_{l,t}\rho_{l',t}}\left(\alpha^\zf_{l,k,t}\alpha^\zf_{l',k,t} \right. \nonumber \\
&\qquad+\left. \E\left\{(\bw_{l,i_t}^\zf)\herm\tilde{\bh}_{l,k}\tilde{\bh}_{l',k}\herm \bw^\zf_{l',i_t}\right\}\right) \nonumber \\ 
&= \sum\limits_{t=1}^{K} \left(\sum\limits_{l=1}^{L}\sqrt{\rho_{l,t}}\alpha^\zf_{l,k,t}\right)^{2} \!\!+ \sum\limits_{t=1}^{K} \sum\limits_{l=1}^{L}\rho_{l,t}(\beta_{l,k}\!-\!\gamma_{l,k}) \nonumber \\
&= (M\!-\!\tp)\!\sum\limits_{t\in\cP_k}\!\left(\sum\limits_{l=1}^{L}\!\sqrt{\rho_{l,t}\gamma_{l,k}}\right)^{2} \!\! +\! \sum\limits_{t=1}^{K}\sum\limits_{l=1}^{L}\rho_{l,t}(\beta_{l,k}\!-\!\gamma_{l,k}),
\end{align}
where in $(a)$ the cross-expectations vanishes as $\tilde{\bh}_{l,k}$ is independent of $\bw^\zf_{l,i_t}$ and zero-mean RV. 
Plugging~\eqref{eq:fzf-num} and~\eqref{eq:fzf-den} into \Eqref{eq:sinr:dl} gives \Eqref{eq:SINR-ZF-alternative}.

\subsection{Proof of Corollary 2}
Plugging~\eqref{eq:PZF-Property} and~\eqref{eq:PZF-MRT-Property}, for $t=k$, into the numerator of~\eqref{eq:sinr-pzf}, and exploiting the independence between the channel estimation errors and the estimates, yields 
\begin{align} \label{eq:num-sinr-pzf}
&\Big| \sum\limits_{l \in \setZ_k} \sqrt{\rho_{l,k}} \EX{(\hat{\bh}_{l,k}+\tilde{\bh}_{l,k})\herm \bw^\pzf_{l,\ik}} \Big. \nonumber \\
&\qquad\;+ \Big.\sum\limits_{p \in \setM_k} \sqrt{\rho_{p,k}} \EX{ (\hat{\bh}_{p,k}+\tilde{\bh}_{p,k})\herm \bw^\mrt_{p,\ik}} \Big|^2 \nonumber \\
&\quad= \left( \sum\limits_{l \in \setZ_k} \sqrt{(M-\tsl)\rho_{l,k}\gamma_{l,k}} + \sum\limits_{p \in \setM_k} \sqrt{M\rho_{p,k}\gamma_{p,k}} \right)^2 \nonumber \\
&\quad= \left( \sum\limits_{l=1}^L \sqrt{(M-\delta_{l,k}\tsl)\rho_{l,k}\gamma_{l,k}}\right)^2, 
\end{align}
where $\delta_{l,k}$ is defined in~\eqref{eq:delta-def}. The first term of the denominator in~\eqref{eq:sinr-pzf} can be decomposed as shown in~\eqref{eq:den-sinr-mrt-original} at the top of the next page.
We first focus on the last term of the RHS in~\eqref{eq:den-sinr-mrt-original}, where $\bw^\mrt_{p,i_t}$ is defined only if UE $t\!\in\!\setW_p$, and consider UE $k\!\in\!\setW_p$. \addtocounter{equation}{1}
\begin{align} \label{eq:mrt-mean-square-combining-channel}
&\sum\limits_{t \in \setW_p \setminus \cP_k} \EX{\left| \sum\limits_{p \in \setM_t} \sqrt{\rho_{p,t}} \bh_{p,k}\herm \bw^\mrt_{p,i_t}  \right|^2} \nonumber \\
&\qquad + \sum\limits_{t \in \cP_k} \EX{\left| \sum\limits_{p \in \setM_t} \sqrt{\rho_{p,t}} \bh_{p,k}\herm \bw^\mrt_{p,i_t}  \right|^2} \nonumber \\
&\stackrel{(a)}{=} \sum\limits_{t \in \setW_p \setminus \cP_k} \sum\limits_{p \in \setM_t} \rho_{p,t} \EX{\left| \bh_{p,k}\herm \bw^\mrt_{p,i_t}  \right|^2} \nonumber \\
&\qquad+ \sum\limits_{t \in \cP_k} \E\left\{ \left| \sum\limits_{p \in \setM_t} \sqrt{\rho_{p,t}} \left(\hat{\bh}_{p,k}\herm + \tilde{\bh}_{p,k}\herm \right) \bw^\mrt_{p,i_t} \right|^2 \right\} \nonumber \\
&\stackrel{(b)}{=} \sum\limits_{t \in \setW_p \setminus \cP_k} \sum\limits_{p \in \setM_t} \rho_{p,t} \EX{\bh\herm_{p,k}\EX{\bw^\mrt_{p,i_t}(\bw^\mrt_{p,i_t})\herm}\bh_{p,k}} \nonumber \\
&\qquad+ \sum\limits_{t \in \cP_k} \E\left\{ \left| \sum\limits_{p \in \setM_t} \sqrt{\rho_{p,t}} \tilde{\bh}_{p,k}\herm \bw^\mrt_{p,i_t} \right|^2 \right\} \nonumber \\
&\qquad +\sum\limits_{t \in \cP_k} \E\left\{ \left| \sum\limits_{p \in \setM_t} \sqrt{\rho_{p,t}} \hat{\bh}_{p,k}\herm \bw^\mrt_{p,i_t} \right|^2 \right\} \nonumber \\
&\stackrel{(c)}{=} \sum\limits_{t \in \setW_p \setminus \cP_k} \sum\limits_{p \in \setM_t} \rho_{p,t} \beta_{p,k} \! + \! \sum\limits_{t \in \cP_k} \sum\limits_{p \in \setM_t} \rho_{p,t} \EX{\left|\hat{\bh}_{p,k}\herm \bw^\mrt_{p,i_t}\right|^2}  \nonumber \\
&\qquad + \sum\limits_{t \in \cP_k} \sum\limits_{p \in \setM_t} \rho_{p,t} \EX{\tilde{\bh}\herm_{p,k}\EX{\bw^\mrt_{p,i_t}(\bw^\mrt_{p,i_t})\herm}\tilde{\bh}_{p,k}} \nonumber \\
&\qquad+ \sum\limits_{t \in \cP_k} \sum\limits_{p \in \setM_t} \sum\limits_{q \in \setM_t \backslash\{p\}} \sqrt{\rho_{p,t}\rho_{q,t}} \EX{\hat{\bh}\herm_{p,k} \bw^\mrt_{p,i_t}} \nonumber \\
&\qquad\qquad \times \EX{\hat{\bh}_{q,k}\herm \bw^\mrt_{q,i_t}} \nonumber \\
&\stackrel{(d)}{=} \sum\limits_{t \in \setW_p \setminus \cP_k} \sum\limits_{p \in \setM_t} \rho_{p,t} \beta_{p,k} + \sum\limits_{t \in \cP_k} \sum\limits_{p \in \setM_t} \rho_{p,t}\beta_{p,k} \nonumber \\
&\qquad + \sum\limits_{t \in \cP_k} \left(  \sum\limits_{p \in \setM_t} \sqrt{M \rho_{p,t}\gamma_{p,k}} \right)^2, 
\end{align}
where we have exploited: in $(a)$, $\bh_{p,k}$ is a zero-mean RV independent of $\bw^\mrt_{p,i_t}$ when $t \in \setW_p\!\setminus\!\cP_k$; in $(b)$, $\tilde{\bh}_{p,k}$ is independent of $\hat{\bh}_{p,k}$ and $\bw^\mrt_{p,i_t}$, when $t \in \cP_k$, and zero-mean RV; in $(c)$, $\hat{\bh}\herm_{p,k} \bw^\mrt_{p,i_t}$ is independent of $\hat{\bh}_{q,k}\herm \bw^\mrt_{q,i_t}$ when $p\!\neq\!q$, as $\hat{\bh}_{l,k}\!\sim\!\CN(\bzero, \gamma_{l,k}\bI_{M})~\forall l,k$. Moreover, $(d)$ follows from 
\begin{align}
\EX{\left|\hat{\bh}_{p,k}\herm \bw^\mrt_{p,i_t}\right|^2} &\stackrel{t \in \cP_k}{=} \EX{\left|\hat{\bh}_{p,k}\herm \bw^\mrt_{p,i_k}\right|^2} \nonumber \\
&\;\;= \frac{1}{M \gamma_{p,k}} \EX{\norm{\hat{\bh}_{p,k}}^4} \nonumber \\
&\;\;= (M+1)\gamma_{p,k},
\end{align}
\begin{align}
&\EX{\hat{\bh}\herm_{p,k} \bw^\mrt_{p,i_t}} \EX{\hat{\bh}_{q,k}\herm \bw^\mrt_{q,i_t}} \nonumber \\
&\quad\stackrel{t \in \cP_k}{=} \frac{1}{M \sqrt{\gamma_{p,k}\gamma_{q,k}}} \EX{\norm{\hat{\bh}_{p,k}}^2} \EX{\norm{\hat{\bh}_{q,k}}^2} \nonumber \\
&\quad\;\;= M \sqrt{\gamma_{p,k}\gamma_{q,k}}.
\end{align}
Conversely, if $k \in \setS_p \implies t \in \setW_p \setminus \cP_k$, and only the first term in~\eqref{eq:mrt-mean-square-combining-channel} remains. Hence,~\eqref{eq:mrt-mean-square-combining-channel} is valid for any UE $k$.

Now, we compute in closed form the first term of the RHS in~\eqref{eq:den-sinr-mrt-original}, where $\bw^\pzf_{l,i_t}$ is defined only if UE $t \in \setS_l$. If UE $k \in \setS_l$, then following the same methodology as in Appendix A, but applying~\eqref{eq:PZF-Property}, yields
\begin{align} \label{eq:den-pzf-k-strong}
&\sum\limits_{t=1}^K \! \EX{\! \left| \sum\limits_{l \in \setZ_t} \sqrt{\rho_{l,t}} \bh_{l,k}\herm \bw^\pzf_{l,i_t} \right|^2\!} \nonumber \\
&\quad= \sum\limits_{t\in\cP_k}\!\left(\sum\limits_{l \in \setZ_t}\sqrt{(M\!-\!\tsl)\rho_{l,t}\gamma_{l,k}}\right)^{2}\! \nonumber \\
&\quad\qquad+ \! \sum\limits_{t=1}^{K}\sum\limits_{l \in \setZ_t}\rho_{l,t}(\beta_{l,k}\!-\!\gamma_{l,k}). 
\end{align}
Conversely, if UE $k \in \setW_l \implies t \in \setS_l \setminus \cP_k$, and, similarly to the MRT case, we have
\begin{align} \label{eq:den-pzf-k-weak}
&\sum\limits_{t=1}^K \! \EX{\! \left| \sum\limits_{l \in \setZ_t} \sqrt{\rho_{l,t}} \bh_{l,k}\herm \bw^\pzf_{l,i_t} \right|^2\!} \nonumber \\
&\quad \! = \! \sum\limits_{t \in \setS_l \setminus \cP_k} \sum\limits_{l \in \setZ_t} \rho_{l,t} \EX{\bh\herm_{l,k}\EX{\bw^\pzf_{l,i_t}(\bw^\pzf_{l,i_t})\herm}\bh_{l,k}} \! \nonumber \\
& \quad = \sum\limits_{t \in \setS_l \setminus \cP_k} \sum\limits_{l \in \setZ_t} \rho_{l,t} \beta_{l,k},
\end{align}
since $\bh_{l,k}$ is a zero-mean RV independent of $\bw^\pzf_{l,i_t}$. Combining~\eqref{eq:den-pzf-k-strong} and~\eqref{eq:den-pzf-k-weak}, for any $k$, we have
\begin{align} \label{eq:den-pzf-mean-square}
&\sum\limits_{t=1}^K \! \EX{\! \left| \sum\limits_{l \in \setZ_t} \sqrt{\rho_{l,t}} \bh_{l,k}\herm \bw^\pzf_{l,i_t} \right|^2\!} \nonumber \\
&\quad=\!\!\sum\limits_{t \in \setS_l \setminus \cP_k}\sum\limits_{l \in \setZ_t}\!\rho_{l,t}(\beta_{l,k}\!-\!\delta_{l,k}\gamma_{l,k}) \!+\! \! \sum\limits_{t\in\cP_k}\sum\limits_{l \in \setZ_t}\!\rho_{l,t}(\beta_{l,k}\!-\!\gamma_{l,k}) \nonumber \\
&\quad\qquad + \sum\limits_{t\in\cP_k}\!\left(\sum\limits_{l \in \setZ_t}\sqrt{(M\!-\!\tsl)\rho_{l,t}\gamma_{l,k}}\right)^{2}\!.
\end{align}

Given $l \in \setZ_t$ and $p \in \setM_t$, $\setZ_t \cap \setM_t = \varnothing$ implies that $\bh_{l,k}\herm \bw^\pzf_{l,i_t}$ and $(\bw^\mrt_{p,i_t})\herm \bh_{p,k}$ are independent, thus the expectation $\EX{\bh_{l,k}\herm \bw^\pzf_{l,i_t} (\bw^\mrt_{p,i_t})\herm \bh_{p,k}}$ in the second term of the RHS in~\eqref{eq:den-sinr-mrt-original} can be divided in two parts, and by using~\eqref{eq:PZF-Property} and~\eqref{eq:PZF-MRT-Property}, it is given by
\begin{align} \label{eq:cross-terms-pzf-mrt}
&\hat{\bh}_{l,k}\herm \bw^\pzf_{l,i_t}\EX{(\bw^\mrt_{p,i_t})\herm \hat{\bh}_{p,k}} \nonumber \\
&\quad= 
\begin{cases}
0, &\text{ if } t \notin \cP_k, \\
\sqrt{M(M-\tsl) \gamma_{l,k}\gamma_{p,k}}, &\text{ if } t \in \cP_k,
\end{cases}
\end{align}
where $t \in \cP_k$ implies that $k$ belongs to the same group of $t$, namely $k \in \setS_l$ and $k \in \setW_p$, and $t \notin \cP_k$ must be intended as $t \in \setS_l \! \setminus \! \cP_k,~t \in \setW_p \! \setminus \! \cP_k$. Plugging~\eqref{eq:cross-terms-pzf-mrt} into the second term of the RHS in~\eqref{eq:den-sinr-mrt-original} gives 
\begin{align} \label{eq:den-cross-terms-pzf-mrt}
&\sum\limits_{t\in\cP_k} \left( 2\sum\limits_{l \in \setZ_t}\sqrt{(M-\tsl)\rho_{l,t}\gamma_{l,k}} \sum\limits_{p \in \setM_t} \sqrt{M\rho_{p,t}\gamma_{p,k}} \right) \nonumber \\ 
&\quad= \sum\limits_{t\in\cP_k} \left[ - \left( \sum\limits_{p \in \setM_t} \sqrt{M\rho_{p,t}\gamma_{p,k}} \right)^2 \right. \nonumber \\
&\quad\quad + \left. \left( \sum\limits_{l \in \setZ_t}\sqrt{(M-\tsl)\rho_{l,t}\gamma_{l,k}}  + \sum\limits_{p \in \setM_t} \sqrt{M\rho_{p,t}\gamma_{p,k}} \right)^2 \right. \nonumber \\
&\quad\quad\left. - \left( \sum\limits_{l \in \setZ_t}\sqrt{(M-\tsl)\rho_{l,t}\gamma_{l,k}} \right)^2 \right],
\end{align}
by using $2XY = (X+Y)^2-X^2-Y^2$.
Lastly, inserting~\eqref{eq:mrt-mean-square-combining-channel},~\eqref{eq:den-pzf-mean-square} and~\eqref{eq:den-cross-terms-pzf-mrt} into~\eqref{eq:den-sinr-mrt-original} yields
\begin{align} \label{eq:den-sinr-pzf}
&\sum\limits_{t = 1}^K \E\left\{ \left| \sum\limits_{l \in \setZ_t} \sqrt{\rho_{l,t}} \bh_{l,k}\herm \bw^\pzf_{l,i_t} + \sum\limits_{p \in \setM_t} \sqrt{\rho_{p,t}} \bh_{p,k}\herm \bw^\mrt_{p,i_t}  \right|^2 \right\} \nonumber \\
&\;=\!\!\sum\limits_{t\notin\cP_k}\!\sum\limits_{l=1}^L\!\rho_{l,t}(\beta_{l,k}\!-\!\delta_{l,t}\delta_{l,k}\gamma_{l,k}) \! +\!\!\! \sum\limits_{t\in\cP_k}\!\sum\limits_{l=1}^L\!\rho_{l,t}(\beta_{l,k}\!-\!\delta_{l,t}\gamma_{l,k}) \nonumber \\
&\quad\quad+\sum\limits_{t\in\cP_k}\!\left(\!\sum\limits_{l=1}^L\!\sqrt{(M\!-\!\delta_{l,t}\tsl)\rho_{l,t}\gamma_{l,k}}\!\right)^2\nonumber \\
&\;\stackrel{(e)}{=} \sum\limits_{t\in\cP_k}\!\left(\!\sum\limits_{l=1}^L\!\sqrt{(M\!-\!\delta_{l,t}\tsl)\rho_{l,t}\gamma_{l,k}}\!\right)^2 \nonumber \\
&\quad\quad+\!\sum\limits_{t=1}^K\!\sum\limits_{l=1}^L\!\rho_{l,t}(\beta_{l,k}\!-\!\delta_{l,t}\delta_{l,k}\gamma_{l,k}),
\end{align}
where in $(e)$ we use the fact that $t \in \cP_k \implies \delta_{l,t} = \delta_{l,k}$ and $\delta_{l,t} = \delta_{l,t}\delta_{l,k}$. Substituting~\eqref{eq:num-sinr-pzf} and~\eqref{eq:den-sinr-pzf} into~\eqref{eq:sinr-pzf} gives~\eqref{eq:sinr-pzf-final}.

\begin{figure*}[!t]
\normalsize
\setcounter{eqcnt6}{\value{equation}}
\setcounter{equation}{49}
\begin{align} \label{eq:den-sinr-mrt-original}
&\sum\limits_{t = 1}^K \E\left\{ \left| \sum\limits_{l \in \setZ_t} \sqrt{\rho_{l,t}} \bh_{l,k}\herm \bw^\pzf_{l,i_t} + \sum\limits_{p \in \setM_t} \sqrt{\rho_{p,t}} \bh_{p,k}\herm \bw^\mrt_{p,i_t}  \right|^2 \right\} =
\sum\limits_{t=1}^K \EX{\left| \sum\limits_{l \in \setZ_t}\sqrt{\rho_{l,t}} \bh_{l,k}\herm \bw^\pzf_{l,i_t} \right|^2} \nonumber \\
&\qquad +2 \sum\limits_{t = 1}^K\mathrm{Re}\left\{ \sum\limits_{l \in \setZ_t}\sum\limits_{p \in \setM_t}\sqrt{\rho_{l,t}\rho_{p,t}} \EX{\bh_{l,k}\herm \bw^\pzf_{l,i_t} (\bw^\mrt_{p,i_t})\herm \bh_{p,k}}\right\}+\sum\limits_{t=1}^K\EX{ \left| \sum\limits_{p \in \setM_t}\sqrt{\rho_{p,t}} \bh_{p,k}\herm \bw^\mrt_{p,i_t}\right|^2\!}
\end{align}
\setcounter{equation}{\value{eqcnt6}}
\hrulefill
\vspace*{4pt}
\end{figure*}

\subsection{Proof of Corollary 3}
The proof of Corollary 3 is almost identical to what is given in Appendix B for Corollary 2. The only difference is that the precoding vector for the MRT at AP $l$ now projects the signal to the $M-\tsl$ dimensional subspace orthogonal to the column space of $\bar{\bH}_l\bESl$. This projection implies that, for any UE $t, k \in\setW_l$, with $t \in \cP_k$
\begin{align}
\EX{\hat{\bh}_{l,k}\herm \bw^\pmrt_{l,i_t}} &= \sqrt{(M-\tsl) \gamma_{l,k}}, \\
\EX{\left|\hat{\bh}_{l,k}\herm \bw^\pmrt_{l,i_t}\right|^2} &\stackrel{(a)}{=}  (M-\tsl+1)\gamma_{l,k},
\end{align}
and
\begin{align}
&\EX{\hat{\bh}_{l,k}\herm \bw^\pmrt_{l,i_t}} \EX{\hat{\bh}_{p,k}\herm \bw^\pmrt_{p,i_t}} \nonumber \\
&\qquad \stackrel{(l \neq p)}{=} \sqrt{(M-\tsl)\gamma_{l,k}(M-\tsp)\gamma_{p,k}},
\end{align}
where $(a)$ follows from~\cite[Lemma 2.9]{Tulino2004}, for a $\tsl \times \tsl$ central complex Wishart matrix with $M$ degrees of freedom satisfying $M \geq \tsl + 1$. While, if $k \in \setS_l$ and $t \in \setW_l \implies t \notin \cP_k$, and
\begin{equation}
\EX{\left|\bh_{l,k}\herm \bw^\pmrt_{l,i_t}\right|^2} = \EX{\left|\tilde{\bh}_{l,k}\herm \bw^\pmrt_{l,i_t}\right|^2} = \beta_{l,k} - \gamma_{l,k},
\end{equation}
since, by design, $\hat{\bh}_{l,k}\herm \bw^\pmrt_{l,i_t} = 0$, and $\tilde{\bh}_{l,k}$ is independent of $\bw^\pmrt_{l,i_t}$.
All other calculations are identical to Appendix B.
 
\bibliographystyle{IEEEtran} 
\bibliography{IEEEabrv,refs}
 
\end{document}